\documentclass[reprint]{revtex4-1}

\usepackage{graphicx}
\usepackage{hyperref}
\usepackage{enumerate}
\usepackage{amsmath,amssymb,dsfont,mathrsfs}

\newcommand{\RR}{\mathds{R}}
\newcommand{\ZZ}{\mathds{Z}}
\newcommand{\EE}{\mathds{E}}
\newcommand{\NN}{\mathds{N}}
\newcommand{\eps}{\varepsilon}
\DeclareRobustCommand{\argmin}{\operatorname*{argmin}}
\DeclareRobustCommand{\argmax}{\operatorname*{argmax}}

\begin{document}

\title{Numerical computation of rare events via large deviation theory}
\author{Tobias Grafke}
\affiliation{Mathematics Institute, University of Warwick, Coventry
  CV4 7AL, United Kingdom}
\email{T.Grafke@warwick.ac.uk}
\author{Eric Vanden-Eijnden}
\affiliation{Courant Institute, New York
  University, 251 Mercer Street, New York, NY 10012, USA}
\email{eve2@cims.nyu.edu}

\date{\today}

\begin{abstract}
  An overview of rare events algorithms based on large deviation
  theory (LDT) is presented. It covers a range of numerical schemes to
  compute the large deviation minimizer in various setups, and
  discusses best practices, common pitfalls, and implementation
  trade-offs. Generalizations, extensions, and improvements of the
  minimum action methods are proposed. These algorithms are tested on
  example problems which illustrate several common difficulties which
  arise e.g.~when the forcing is degenerate or multiplicative, or the
  systems are infinite-dimensional. Generalizations to processes
  driven by non-Gaussian noises or random initial data and parameters
  are also discussed, along with the connection between the LDT-based
  approach reviewed here and other methods, such as stochastic field
  theory and optimal control. Finally, the integration of this
  approach in importance sampling methods using e.g.~genealogical
  algorithms is explored.
\end{abstract}

\maketitle

\tableofcontents

\newpage

\begin{quote}
  \textit{Rare events often have a drastic impact despite their low frequency
  of occurrence. Examples include hurricanes, financial crises, heat
  waves, or tsunamis, that are few and far between but have
  devastating consequences. Other important phenomena such as phase
  transitions, chemical reactions, or conformational changes of
  biomolecules also involve rare events. The accurate description of
  these events is complicated, since their low rate of occurrence
  makes them hard to observe both in experiments and in
  simulations. In many cases, when a rare event occurs, it does so in
  its least unlikely form, the instanton, rendering all other
  realizations of the same event negligible in comparison. Whenever
  such a situation holds, a large deviation principle (LDP) quantifies
  this concentration phenomenon. The LDP specifies a deterministic
  optimization problem to identify the instanton, and allows the
  estimation of its probability. In this review, we discuss numerical
  algorithms to solve the large deviation optimization problem,
  compare their associated trade-offs, and present best practices,
  pitfalls, improvements, and generalizations.}
\end{quote}

\section{Introduction}
\label{sec:introduction}

Rare but important events are by definition difficult to observe, both
in experiments and in simulations. In order to design efficient
schemes for the numerical computation of these events one therefore
typically resorts to one of the following two strategies: either
manipulation of the system in a controlled way that makes rare events
more likely and can be corrected \textit{a~posteriori}; or computation
of a single dominant event characterizing the possible ways the rare
event happens. The first approach can be categorized as importance
sampling; the second can be justified within large deviation theory
(LDT) and leads to an action minimization problem to be solved. In
this review, we focus mainly on algorithmic developments in the second
class, and discuss the interplay between this LDT-based approach and
importance sampling towards the end of our paper.  Specifically, in
section~\ref{sec:rare-event-algor} we discuss rare event algorithms
based on the global minimization of LDT action functionals, suitable
for computing paths by which infrequent transitions between two
prescribed states occur. Subsequently, in
section~\ref{sec:rare-event-algor-1}, we explain how to calculate
large deviation minimizers in the context of the estimation of rare
expectations dominated by tail statistics. In
section~\ref{sec:gener-non-gauss}, we generalize these two approaches
to the non-Gaussian case. In section~\ref{sec:systems-with-random}, we
demonstrate a generalization to arbitrary dynamical systems with
random initial conditions and
parameters. Section~\ref{sec:usage-inst-other} suggests possibilities
to use the minimizing trajectories obtained by the earlier algorithms
as input for importance sampling algorithms. Finally, some concluding
remarks are presented in section~\ref{sec:conclusion}.

In the reminder of the present section we review the aspects of LDT
relevant to our purpose, with focus on Freidlin-Wentzell theory for
dynamical systems subject to random noise of low amplitude.

\subsection{Freidlin-Wentzell theory of large deviations}
\label{sec:freidl-wentz-theory}

Consider a dynamical system with variables $X_t^\eps$ in $\RR^d$,
subject to small random perturbations that are additive Gaussian and
white in time. Assuming that the noise amplitude scales with the
smallness-parameter $\eps$, the evolution of the stochastic variables
$X_t^\eps$ is described by the stochastic differential equation (SDE)
\begin{equation}
  \label{eq:1}
  dX_t^\eps = b(X_t^\eps)\,dt + \sqrt{\eps} \sigma\,dW_t\,,
  \quad X^\eps_0=x,\ \ t\ge0\,,
\end{equation}
with deterministic drift $b:\RR^d\to\RR^d$, noise covariance
$a=\sigma\sigma^\top$ with $\sigma\in\RR^{d\times d}$, and where $W_t$
is a $d$-dimensional Wiener process -- for simplicity, we assume that
$\sigma$ is independent of the system's position (i.e.~the noise is
additive): the generalization to multiplicative noise is
straightforward and will be discussed through examples. We are
interested in situations where the stochastic process~(\ref{eq:1})
realizes a certain event, for example when the trajectory ends at time
$T$ in a given set $A\subset\RR^d$, so that $X_T^\eps\in A$. Even if
these events are impossible in the deterministic system ($\eps=0$),
they will in general occur in the presence of noise ($\eps>0$) but
they become rarer and rarer in the low noise limit, $\eps\to0$.

Large deviation theory (LDT) gives a precise characterization of this
decay of probability: The probability of observing any sample path
close to a given function $\phi(t)$ can be estimated as
\begin{equation}
  \label{eq:2}
  P\big\{\sup_{t\in[0,T]} \|X_t^\eps-\phi(t)\|<\delta\big\} \asymp \exp\left(-\eps^{-1} S_T(\phi)\right)\,,
\end{equation}
for small enough $\delta>0$, where $\asymp$ denotes log-asymptotic
equivalence (i.e.~for $\eps\to0$, the ratio of the logarithms of both
sides converges to 1). The functional $S_T(\phi)$ is called the
\emph{rate function} or \emph{action functional}, and it is generally given by
\begin{equation}
  \label{eq:action}
  S_T(\phi)=
  \begin{cases}
    \int_0^T L(\phi,\dot\phi)\,dt & \text{if the integral converges,}\\
    \infty & \text{otherwise.}
  \end{cases}
\end{equation}
Here we defined the \emph{Lagrangian} $L(\phi,\dot\phi)$, which for
the concrete example of equation~(\ref{eq:1}) is given by:
\begin{equation}
  L(\phi,\dot\phi) = \tfrac12 \|\dot\phi - b(\phi)\|_a^2
\end{equation}
via the $a$-metric induced by the inner product
$\|v\|_a^2=\langle v,a^{-1}v\rangle$ (assuming invertibility of $a$ for
simplicity).

The probability of observing the event $X_T^\eps\in A$ consists of
contributions of the sample paths close to all the possible
$\phi(t)\in\mathscr C=\{\phi(t)\in C([0,T],\RR^d)\ |\ \phi(0)=x,
\phi(T)\in A\}$, and each of these contributions scales according to
equation~(\ref{eq:2}). Consequently, in the limit $\eps\to0$, the only
contribution that matters is that coming from the trajectory
$\phi^*(t)$ with the smallest action $S_T(\phi^*)$. We call
\begin{equation}
  \label{eq:3}
  \phi^*(t) = \argmin_{\phi\in\mathscr C} S_T(\phi)
\end{equation}
the \emph{maximum likelihood pathway} (MLP) or \emph{instanton}. It
constitutes the \emph{least unlikely} trajectory to realize the rare
event, in the sense that almost surely all sample paths conditioned on
the rare event will be arbitrarily close to $\phi^*(t)$. More
precisely, for all $\delta>0$ sufficiently small, we have
\begin{equation}
  \lim_{\eps\to0} P\big\{\sup_{t\in[0,T]} \|X_t^\eps-\phi^*(t)\|
  <\delta\ \big|\  X_T^\eps\in A\big\} = 1\,.
\end{equation}
The efficient numerical solution of the minimization problem~(\ref{eq:3}) for
different rare events (and therefore different sets of trajectories
$\mathscr C$ to minimize over) lies at the core of this work.

If the action at the instanton, $S_T(\phi^*)$ is zero, the
corresponding trajectory fulfills $\dot \phi = b(\phi)$ and can be
considered \emph{deterministic}, i.e.~the corresponding evolution is
the one selected by the deterministic dynamics ($\eps=0$). If, on the
other hand, the action at the instanton is finite, the probability of
observing the corresponding event in a given time frame $T$ decays to
zero as indicated by equation~(\ref{eq:2}).

LDT additionally permits the analysis of the effect of infinitesimal
perturbations over an infinite time interval, $T\to\infty$, on which
these rare events almost surely happen. The central object in this
context is the \emph{quasipotential}, defined as
\begin{equation}
  \label{eq:quasipotential}
  V(x,y) = \inf_{T>0}\min_{\phi\in\mathscr C_{x,y}} S_T(\phi)\,,
\end{equation}
where
$\mathscr C_{x,y} = \{\phi\in C([0,T],\RR^d)\ \|\ \phi(0)=x,
\phi(T)=y\}$. The quasipotential characterizes the long time behavior
of the system. For example, if the deterministic system
$\dot X = b(X)$ possesses only one single stable fixed point $\bar x$,
with basin of attraction $\RR^d$, then the density $\rho(x)$
associated with the invariant measure of equation~(\ref{eq:1}) can be
written in the limit $\eps\to0$ as
\begin{equation}
  \rho(x) \asymp \exp\left(-\eps^{-1} V(\bar x, x)\right)\,.
\end{equation}
Similarly, in situations where the deterministic system has multiple
fixed points $\bar x_i$, the mean first passage time $\tau_{i,j}$
between the  basins of attraction of  neighboring $\bar x_i$ and $\bar x_j$,
\begin{equation}
  \tau_{i,j} = \EE \inf \{ t>0\ |\ X(0)=\bar x_i, \|X(t)-\bar x_j\|<\delta\}
\end{equation}
with $\delta>0$ small enough, can be estimated in the small noise
limit as
\begin{equation}
  \ \tau_{i,j} \asymp \exp\left(\eps^{-1} V(\bar x_i, \bar x_j)\right)\,.
\end{equation}
This result also allows the investigation of the long time dynamics of
the system by mapping it onto a Markov jump process whose states are
the fixed points $\bar x_i$, $\bar x_j$, etc. and whose transition
rates are $k_{i,j} \asymp \tau_{i,j}^{-1}$.


All these examples show that it is useful to have access to the
minimizing trajectory $\phi^*$ to describe rare events: First it gives
the typical way a rare event is observed, enabling us to identify
their mechanism. Second it allows the estimation of their probability
of occurrence, and their expected recurrence time. Third it gives the
relative probability of multiple typical (i.e.~deterministically
stable) states, and the most likely way by witch transitions between
them occur.

\subsection{Hamiltonian principle and connections to classical
  mechanics and field theory}

The minimization problem in equation~(\ref{eq:3}) to find the
instanton precisely corresponds to Hamilton's principle from classical
mechanics, $\delta S_T(\phi)/\delta \phi=0$. As a consequence, the
methods and ideas from classical mechanics are transferable to our
situation. In particular, the variational problem can be solved by
seeking solutions of the corresponding Euler-Lagrange equation,
\begin{equation}
  \frac{\partial L}{\partial \phi} - \frac{d}{dt}\frac{\partial L}{\partial \dot\phi}=0\,,
\end{equation}
which, for a system of the type~(\ref{eq:1}), gives
\begin{equation}
  \label{eq:4}
  a^{-1}\ddot \phi + \left(a^{-1}\nabla b(\phi) - \nabla b(\phi)^\top
    a^{-1}\right) \dot\phi
  + \nabla \langle b(\phi),a^{-1} b(\phi)\rangle = 0\,.
\end{equation}
Several algorithms presented below aim at the numerical
solution of the second order equation~(\ref{eq:4}).

Similarly inspired by classical mechanics, we can define a
\emph{conjugate momentum}
\begin{equation}
  \theta = \frac{\partial L(\phi,\dot\phi)}{\partial \dot\phi}\,,
\end{equation}
and a \emph{Hamiltonian} as Fenchel-Legendre transform of the
Lagrangian,
\begin{equation}
  H(\phi,\theta) = \sup_{y}\left(\langle \theta,y\rangle - L(\phi,y)\right)\,,
\end{equation}
such that, assuming convexity of $L(\phi,\dot\phi)$ in
$\dot\phi$,
\begin{equation}
  L(\phi,\dot\phi) = \sup_{\theta} \left(\langle\dot\phi,\theta\rangle - H(\phi,\theta)\right)\,.
\end{equation}
The minimization~(\ref{eq:3}) is then equivalent to solving
Hamilton's equations of motion, or \emph{instanton equations},
\begin{equation}
  \label{eq:5}
  \begin{cases}
    \dot\phi = \nabla_\theta H(\phi,\theta)\\
    \dot\theta = -\nabla_\phi H(\phi,\theta)\,.
  \end{cases}
\end{equation}
For the system~(\ref{eq:1}), the Hamiltonian is given by
\begin{equation}
  \label{eq:Hamiltonian-SDE}
  H(\phi,\theta) = \langle b(\phi),\theta\rangle
  + \tfrac12 \langle \theta, a \theta\rangle\,,
\end{equation}
so that the instanton equations read
\begin{equation}
  \label{eq:6}
  \begin{cases}
    \dot\phi = b(\phi) + a \theta\\
    \dot\theta = -(\nabla b(\phi))^\top \theta\,.
  \end{cases}
\end{equation}
Solving the instanton equations~(\ref{eq:5}) constitutes another
possible approach to solving the minimization problem~(\ref{eq:3}),
but care has to be taken to obtain the correct boundary conditions
for~(\ref{eq:5}), depending on the rare event under
consideration---this point will be discussed at length below and we
will see that these boundary conditions make working with \eqref{eq:4}
more appropriate in some cases and with \eqref{eq:6} in others.

The Hamiltonian $H(\phi,\theta)$ is a conserved quantity along the
minimizing trajectory, since
\begin{equation}
  dH/dt = \langle \nabla_\phi H, \dot
  \phi\rangle + \langle\nabla_\theta H ,\dot\theta\rangle= 0\,.
\end{equation}
Additional simplifications apply in the special case that the
minimizing trajectory starts at rest at a fixed point $\bar x$ of the
deterministic dynamics, in which case individually $b(\bar x)=0$ and
$\theta=0$. This necessitates at the same time that the transition
time $T$ diverges to $\infty$, and furthermore that the Hamiltonian
vanishes, $H(\phi,\theta)=0$. This property can in turn be used to
rewrite the action functional~(\ref{eq:action}) as
\begin{equation}
  \label{eq:maupertuis}
  S(\phi) \!=\! \int_0^\infty L(\phi,\dot\phi)\,dt \!=\! \int_0^\infty
  (\langle \dot\phi,\theta\rangle - H(\phi,\theta))\,dt \!=\! \int \langle
  \theta, d\phi\rangle\,.
\end{equation}
Writing the action in this form is known as the Maupertuis principle
in mechanics, and it offers an approach at solving the double
minimization problem to calculate the quasipotential
in~\eqref{eq:quasipotential}.


Interestingly, there is a parallel between the LDT discussed above
and concepts from field theory applied to stochastic
systems. Specifically the Janssen-de~Dominicis
formalism~\cite{janssen:1976, dominicis:1976} based on the
Martin-Siggia-Rose path integral~\cite{martin-siggia-rose:1973}
considers computing expectations as path-integrals over all possible
noise realizations, and performs a change of variables to the field
variable itself. The constraint of the dynamics is embedded as
Lagrange multiplier, which gives rise to an additional \emph{auxiliary
  field}, corresponding to the conjugate momentum. Similarly, the
minimization problem~(\ref{eq:3}) then amounts to finding a
semiclassical trajectory as saddle-point approximation of the action
functional. It is this correspondence which is the root of the terms
``action functional'' and ``instanton'' for the rate function and its
minimizer.  Noteworthy in this context is also the Doi-Peliti
formalism \cite{doi:1976,peliti:1986}, which follows a similar route
for dominant reaction pathways.

\subsection{Detailed balance and gradient flows}

A special case of interest is when the dynamics is a \emph{diffusion
  in a potential},  with the drift given by the negative gradient
of a potential $U:\RR^d\to\RR$ and $\sigma=\sqrt{2}\text{Id}$, such
that equation~(\ref{eq:1}) becomes
\begin{equation}
  \label{eq:potential-diffusion}
  dX_t^\eps = -\nabla U(X_t^\eps)\,dt + \sqrt{2\eps}\,dW_t\,.
\end{equation}
Suppose that we look at the calculation of the
quasipotential~\eqref{eq:quasipotential} between two local minima of
$U$ located at $x_a$ and $x_b$ and with adjacent basins of attraction.
In this case, the minimum of \eqref{eq:quasipotential} is approached
by taking either $\dot\phi=-\nabla U(\phi)$ (in which case the action
is zero), or we realize that
\begin{align*}
  S_T(\phi) &= \tfrac14 \int_0^T | \dot\phi + \nabla U(\phi)|^2\,dt\\
  &= \int_0^T |\dot\phi - \nabla U(\phi)|^2\,dt + \int_0^T \langle \dot\phi,\nabla U(\phi)\rangle\,dt\\
  &= \int_0^T |\dot\phi - \nabla U(\phi)|^2\,dt + \phi(T)-\phi(0)\,.
\end{align*}
Now, since the last terms depend only on the trajectory end-points, we
are free to choose $\dot\phi = \nabla U(\phi)$ to make the first
integral disappear. As a consequence, for a diffusion in a potential
landscape of the form~(\ref{eq:potential-diffusion}), to calculate the
minimum involved in the quasipotential, we can patch together the
solutions of $\dot\phi = \pm \nabla U(\phi)$ that connects $x_a$ and
$x_b$ via the saddle point $x_s$ of minimum potential. We can
interpret that to say that the minimum is achieved by following either the
deterministic dynamics $\dot\phi=-\nabla U(\phi)$, or its
\emph{time-reversed} version. This is nothing but a manifestation of
\emph{time-reversal symmetry} that is the consequence of the random
process defined by equation~(\ref{eq:potential-diffusion}) being in
\emph{detailed balance}. 

This simple relationship between the tangential vector $\dot\phi$ and
the deterministic drift $\nabla U$ simplifies the computation of the
minimizers significantly. In particular, we realize that minimizers
for gradient flows are \emph{heteroclinic orbits} of the deterministic
drift. As such, they are numerically accessible by the \emph{string
  method}~\cite{e-ren-vanden-eijnden:2002,e-ren-vanden-eijnden:2007}.

Similar simplifications as the above can be realized for any system in
detailed balance, and as a results, its large deviation minimizers are
always heteroclinic orbits of the associated \emph{generalized}
gradient flow (but not necessarily of a traditional gradient
flow)~\cite{grafke:2018}.

\section{Rare event algorithms for noise-induced transitions}
\label{sec:rare-event-algor}

In this section, we want to consider a particular sub-class of
problems of the form discussed in
section~\ref{sec:freidl-wentz-theory}: The computation of the optimal
noise-induced transition trajectory from one basin of attraction of
the deterministic dynamics to a neighboring one. In this form, the
minimization problem~(\ref{eq:3}) becomes
\begin{equation}
  \phi^*(t) = \argmin_{\phi\in\mathscr C_{x,y}} S_T(\phi)\,,
\end{equation}
where
$\mathscr C_{x,y} = \{\phi\in C([0,T],\RR^d)\ |\ \phi(0)=x,
\phi(T)=y\}$, and the instanton constitutes the maximum likelihood
transition trajectory between the two deterministically stable fixed
points $x$ and $y$. By additionally minimizing over the transition
time $T$, the resulting instanton can be used to compute the
quasipotential~(\ref{eq:quasipotential}). This calculation can be
performed by applying the minimum action
method~\cite{e-ren-vanden-eijnden:2004}, that discretizes the action
functional~(\ref{eq:action}), and considers the discrete (finite
dimensional) gradient as descent direction for numerical minimization
algorithms, such as gradient descent or quasi-Newton methods. In
section~\ref{sec:simpl-geom-minim}, we present a simplified version
of the minimum action method, and discuss its implementation
details. In section~\ref{sec:exampl-metast-simple}, this method is
then illustrated by applying it to a simplified metastable climate
model. Finally in section~\ref{sec:inst-stoch-part} we discuss
generalizations to stochastic partial differential equations, and
consider the example of the stochastic Burgers-Huxley model in
section~\ref{sec:exampl-stoch-burg}.

\subsection{A simplified geometric minimum action method}
\label{sec:simpl-geom-minim}

One obvious disadvantage of a straightforward discretization of the
Freidlin-Wentzell action functional is its inability to treat infinite
transition times. In the context of the quasipotential, we are looking
for transition trajectories of arbitrary transition time $T$, which
generally diverges, $T\to\infty$, since  the trajectory contains fixed
points. The minimum of the outer minimization in the computation of the
quasipotential,
\begin{equation}
  V(x,y) = \inf_{T>0}\min_{\phi\in\mathscr C_{x,y}} S_T(\phi)\,,
\end{equation}
is simply not attained for any finite $T$ in these cases. This
complication was successfully addressed with the \emph{geometric}
minimum action method~\cite{heymann-vanden-eijnden:2008}, which
instead considers a minimization over the space of arc-length
parametrized curves that may remain finite even for diverging
transition time. In this section, we want to introduce a simplified
version of this geometric picture, allowing us to formulate an
algorithm to compute of the geometric minimizer with a lower number of
derivatives of the Hamiltonian.

Based on the Maupertuis principle~(\ref{eq:maupertuis}), the
minimizing trajectory $\phi$ between two fixed points $x$ and $y$ for
arbitrary transition time $T$ fulfills $H(\phi, \theta)=0$, and the
corresponding action~(\ref{eq:maupertuis}) is given by
\begin{equation}
  \label{eq:S1}
  S(\phi) = \int_x^y\langle\theta,d\phi\rangle\,.
\end{equation}
This form of the action makes it obvious that the action is invariant
under reparametrization: The total action is a line-integral along the
minimizer, and we are free to choose any parametrization to describe
it. This enables us to treat infinite time-intervals with finitely
many discretization points, for example by parametrizing (\ref{eq:S1})
by normalized arc-length.

The minimization problem~(\ref{eq:3}) can be rewritten as a nested
optimization problem,
\begin{equation}
  \label{eq:full-sgmam}
  \phi^* = \argmin_{\phi\in\mathscr C_{x,y}} \sup_{\theta: H(\phi,\theta)=0} E(\phi,\theta)\,,
\end{equation}
with
\begin{equation}
  \label{eq:e-integral}
  E(\phi,\theta) = \int_0^1 \langle \theta, \phi'\rangle\,ds\,.
\end{equation}
Here the prime denotes differentiation with respect to the
parametrization $s$ we choose for $\phi$ and $\theta$, and we impose
$\|\phi'\|_\sim=L=\text{const}$, with $L$ the length of the
curve. Note that the algorithm works independently of the choice of
the norm, and we will discuss appropriate norms at the end of this
section. Therefore, in the following, the norm $\|\cdot\|_\sim$ and
corresponding inner product $\langle\cdot,\cdot\rangle_\sim$ are to be
seen as a placeholder for our preferred choice.

Let
\begin{equation}
  \label{eq:e-problem}
  E^*(\phi) = \sup_{\theta:H(\phi,\theta)=0} E(\phi,\theta)
\end{equation}
and $\theta^*(\phi)$ be the solution of the inner optimization
problem~(\ref{eq:e-problem}), such that
$E(\phi,\theta^*(\phi))=E^*(\phi)$. Then, equivalently, $\theta^*(\phi)$
fulfills the Euler-Lagrange equation for the constrained maximization
problem~(\ref{eq:e-problem}). Using $\delta E(\phi,\theta)/\delta
\theta = \phi'$, this Euler-Lagrange equation reads
\begin{equation}
  \label{eq:contrained-sgmam}
  \phi' = \mu \nabla_\theta
    H(\phi,\theta)\,,
\end{equation}
where $\mu(s)$, $s\in[0,1]$, is a Lagrange multiplier to enforce
the constraint of a vanishing Hamiltonian. This Lagrange multiplier is
explicitly computable by multiplying
equation~(\ref{eq:contrained-sgmam}) by $\phi'$ and solving
for $\mu$, i.e.
\begin{equation}
  \mu = \frac{\|\phi'\|_\sim^2}{\langle \phi',\nabla_\theta H\rangle_\sim}\,.
\end{equation}
Similarly, using $\delta E(\phi,\phi)/\delta \phi = -\theta'$ the
functional derivative of $E^*(\phi)$ with respect to $\phi$ can be
expressed as
\begin{align}
  \frac{\delta E^*(\phi)}{\delta \phi} 
  &= -{\theta^*}'(\phi) + \mu \nabla_\theta H(\phi,\theta^*(\phi)) \nabla_\phi \theta^*(\phi)\nonumber\\
  &= -{\theta^*}'(\phi) - \mu \nabla_\phi H(\phi,\theta^*(\phi))\,,
\end{align}
where the last step makes use of
\begin{equation*}
  \nabla_\phi H(\phi,\theta^*(\phi)) = -\nabla_\theta H(\phi, \theta^*(\phi)) \nabla_\phi \theta^*(\phi)\,,
\end{equation*}
which holds by definition due to $H(\phi,\theta^*(\phi))=0$.

Note how in this formulation the reparametrization into arc-length
emerges naturally as Lagrange multiplier $\mu$ to enforce the
Hamiltonian constraint. In particular, comparing
equation~(\ref{eq:contrained-sgmam}) with Hamilton's equation with
respect to physical time, $d\phi/dt = \nabla_\theta H(\phi,\theta)$
shows that the Lagrange multiplier $\mu$ is nothing but the change of
parametrization, $\mu = dt/ds$ from physical time to arc-length
parametrization.

Taking these equivalences, the nested optimization
problem~(\ref{eq:full-sgmam}) can now be solved in an iterative
manner. Starting from a the $k$-th guess $\phi^k$ for the transition
trajectory,
\begin{enumerate}[(i)]
\item solve the inner constrained optimization problem
  \begin{equation*}
    \theta^k = \theta^*(\phi^k) = \argmax_{\theta: H(\phi^k,\theta)=0} E(\phi^k,\theta)\,,
  \end{equation*}
\item compute a descent direction for the outer optimization
  problem,
  \begin{equation}
    \label{eq:sgmam-dk-update}
    d^k = \delta E^*(\phi^k)/\delta\phi^k
    = \dot\theta^k + \nabla_\phi H(\phi^k,\theta^k)\,,
  \end{equation}
\item descent along the descent direction, for example by gradient
  descent, pre-conditioned with $\mu^{-1}$, and step-length $\alpha$,
  \begin{equation}
    \label{eq:sgmam-phi-update}
    \phi^{k+1} = \phi^k + \alpha\mu^{-1} d^k\,,
  \end{equation}
  to obtain the next guess $\phi^{k+1}$, and finally
\item iterate until convergence.
\end{enumerate}

For the specific case of the small-noise Gaussian SDE,
equation~(\ref{eq:1}), this algorithm can be even more simplified. In
particular, the inner constrained optimization problem to find
$\theta^*(\phi)$ can be solved analytically, instead of relying on
numerical optimization. Taking the Euler-Lagrange
equation~(\ref{eq:contrained-sgmam}) for the inner optimization
problem, together with the specific form of the
Hamiltonian~(\ref{eq:Hamiltonian-SDE}), yields
\begin{equation*}
  \phi' = \mu \nabla_\theta H(\phi,\theta^*(\phi)) = \mu (b(\phi) + a\theta^*(\phi))\,,
\end{equation*}
so that
\begin{equation}
  \label{eq:theta-SDE}
  \theta^*(\phi) = a^{-1}(\mu^{-1}\phi' - b(\phi))\,.
\end{equation}
On the other hand, the Lagrange multiplier $\mu$ is directly available
without knowledge of $\theta^*$: Since
\begin{equation}
  \|\nabla_\theta H\|_a^2 \!=\! \|b+a\theta\|_a^2 \!=\! \|b\|_a^2 + 2\langle b,\theta\rangle + \|a\theta\|_a^2 \!=\! \|b\|_a^2 + 2H \!=\! \|b\|_a^2\,,
\end{equation}
we conclude that
\begin{equation}
  \label{eq:mu-SDE}
  \mu = \frac{\|\nabla_\theta E(\phi,\theta^*(\phi))\|_a}{\|\nabla_\phi H(\phi,\theta^*(\phi))\|_a} = \frac{\|\phi'\|_a}{\|b+\theta\|_a} = \frac{\|\phi'\|_a}{\|b\|_a}\,.
\end{equation}
(\ref{eq:mu-SDE}) naturally leads to the choice
$\|\cdot\|_\sim = \|\cdot\|_a$. The descent direction is then
immediately available as
\begin{equation*}
  d^k = {\theta^k}' + (\nabla b(\phi^k))^\top \theta^k\,,
\end{equation*}
with $\theta^k$ and $\mu$ given by (\ref{eq:theta-SDE}) and (\ref{eq:mu-SDE}), respectively.

We want to make a few points about possible pitfalls and best practices.
\begin{itemize}
\item Even though any parametrization $s(t)$ is permissible, as
  discussed above it is natural to choose arc-length, such that
  $\|\phi'\|_\sim=\text{const}$. This parametrization can be enforced,
  as in the improved string method~\cite{e-ren-vanden-eijnden:2007}
  and the original geometric minimum action
  method~\cite{heymann-vanden-eijnden:2008}, by interpolation along
  the trajectory. This avoids stiff terms enforcing the
  parametrization constraint.
\item Pre-conditioning is necessary to obtain good
  convergence. Pre-conditioning with $\mu^{-1}$ is necessary to ensure
  convergence around fixed points. Additionally, pre-conditioning with
  $\nabla_\theta \nabla_\theta H$ is often beneficial. This
  corresponds to the noise covariance $a$ in the SDE case. For
  general Hamiltonians, this comes at the cost of needing to compute
  higher derivatives of the Hamiltonian, which one might want to
  avoid. Details about these considerations are discussed
  in~\cite{grafke-schaefer-vanden-eijnden:2017}.
\item The choice of norm has to be taken with care as well. For the
  additive Gaussian case, as pointed out above, it is natural to use
  $\|\cdot\|_\sim = \|\cdot\|_a$. This generalizes to $\langle \cdot,
  (\nabla_\theta \nabla_\theta H)^{-1} \cdot\rangle$, which is the
  choice of the traditional gMAM. For simplicity, the Euclidean norm
  might be preferred in the general case to avoid the computation of
  higher order derivatives of $H$.
\end{itemize}

\subsection{Example: Metastability in a simple climate model}
\label{sec:exampl-metast-simple}

\begin{figure*}
  \begin{center}
    \includegraphics[width=0.49\textwidth]{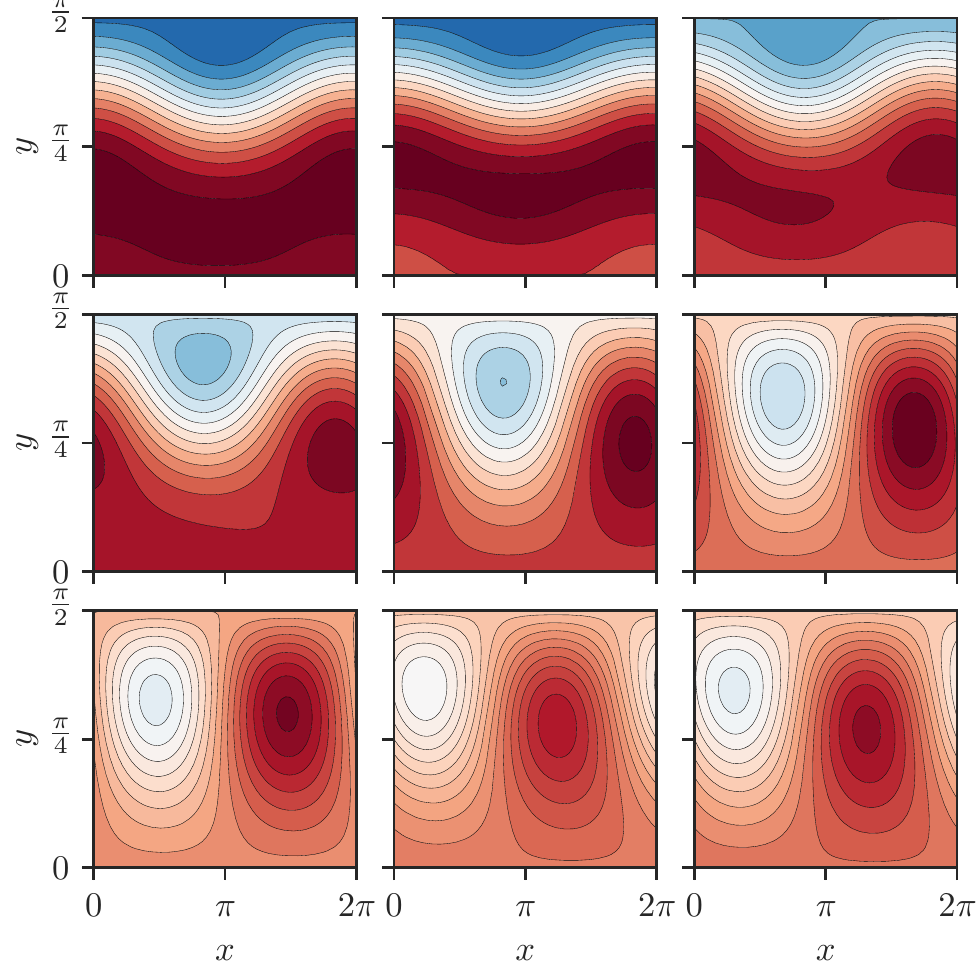}
    \includegraphics[width=0.49\textwidth]{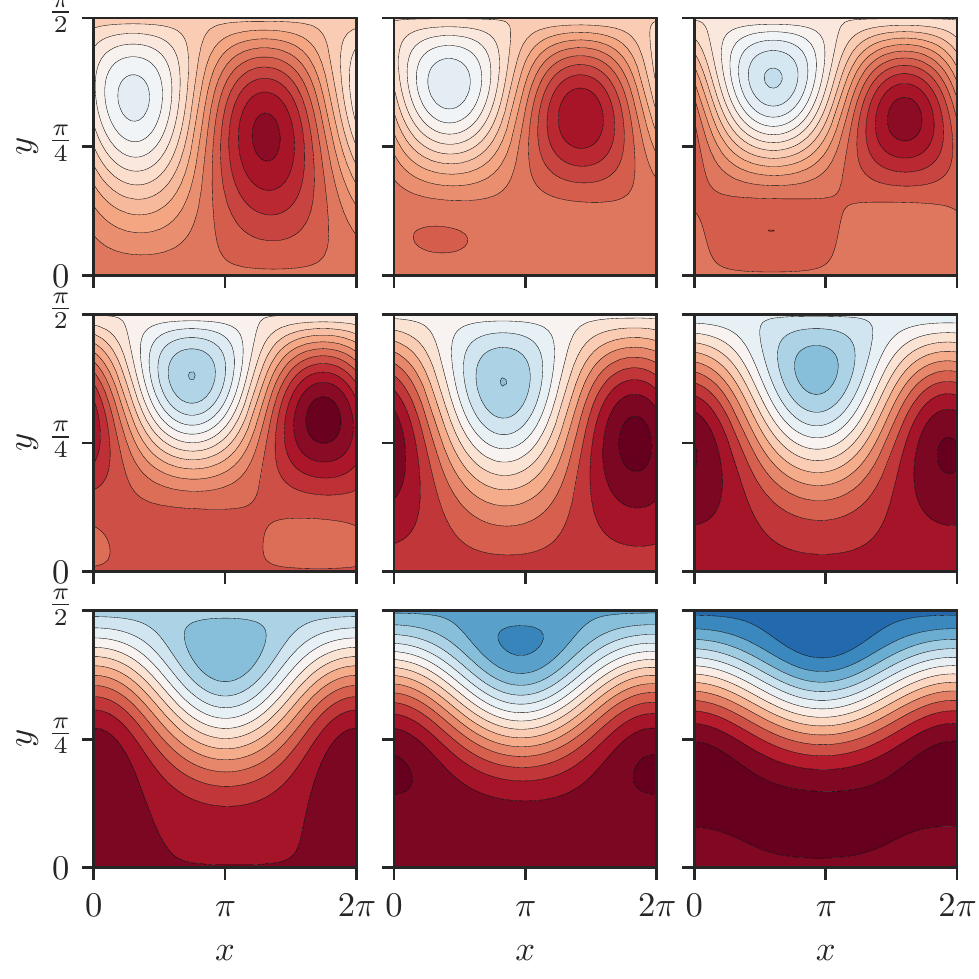}
  \end{center}
  \caption{Left: Stream function $\psi(x,y)$ along the transition from
    zonal to blocked configuration, where the arclength-parameter is
    increased in lexicographic order. The central configuration is the
    unstable saddle configuration on the separatrix between the basins
    of attraction of zonal and blocked configuration. Right: Stream
    function $\psi(x,y)$ along the transition from blocked to zonal
    configuration. Notably, this backward transition is not identical
    to the time-reversal of the forward transition depicted on the
    left, but again the same saddle is visited, as visible in the
    center field.\label{fig:cdv-transition}}
\end{figure*}
\begin{figure}
  \begin{center}
    \includegraphics[width=200pt]{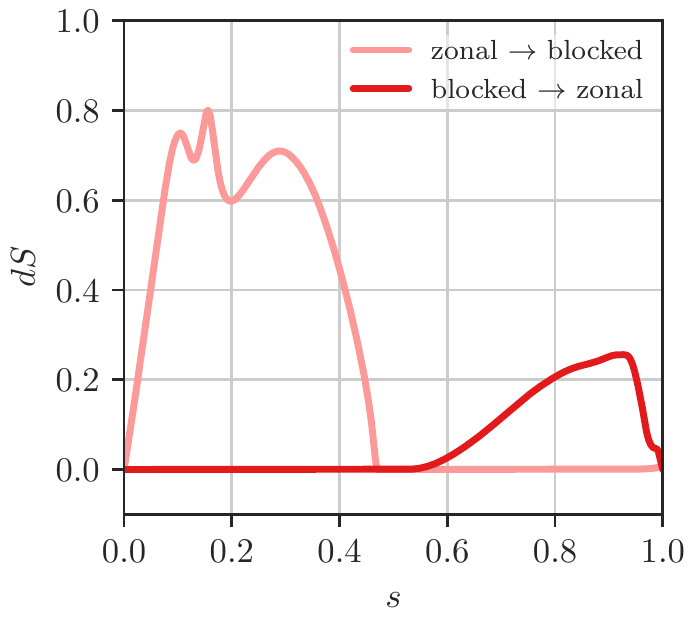}
  \end{center}
  \caption{Action density along the transition trajectories between
    the zonal and the blocked configuration. As clearly visible, the
    transition towards the blocked state occurs at higher action,
    making the blocked state relatively more
    stable.\label{fig:cdv-action}}
\end{figure}

We want to demonstrate the effectiveness of the algorithm introduced
in section~\ref{sec:simpl-geom-minim} by applying it to a problem
motivated by meta-stability in a simplified climate model introduced
by Charney and DeVore~\cite{charney-devore:1979}. Starting from the
two-dimensional barotropic vorticity equation for the atmospheric
flow, a projection on the 6 dominant Fourier modes is performed,
resulting in an SDE system
\begin{equation}
  \label{eq:cdv}
  \begin{split}
    d x_1 &= \left(\tilde \gamma_1 x_3 - C(x_1-x_1^*)\right) dt + \sqrt{2\epsilon} dW_1\,,\\
    d x_2 &= \left(-(\alpha_1x_1 - \beta_1)x_3 - Cx_2 -\delta_1 x_4x_6\right) dt + \sqrt{2\epsilon} dW_2\,,\\
    d x_3 &= \left((\alpha_1x_1-\beta_1)x_2 - \gamma_1 x_1 - Cx_3 + \delta_1x_4x_5\right) dt + \sqrt{2\epsilon} dW_3\,,\\
    d x_4 &= \left(\tilde\gamma_2 x_6-C(x_4-x_4^*)+\eta(x_2x_6-x_3x_5)\right) dt + \sqrt{2\epsilon} dW_4\,,\\
    d x_5 &= \left(-(\alpha_2x_1-\beta_2)x_6-Cx_5-\delta_2x_3x_4\right) dt + \sqrt{2\epsilon} dW_5\,,\\
    d x_6 &= \left((\alpha_2x_1-\beta_2)x_5 - \gamma_2x_4 - Cx_6 + \delta_2x_2x_4\right) dt + \sqrt{2\epsilon} dW_6\,,
  \end{split}
\end{equation}
where, for $m\in\{1,2\}$,
\begin{equation}
  \begin{split}
    \alpha_m &= \frac{8\sqrt{2}}{\pi} \frac{m^2}{4m^2-1}\frac{b^2+m^2-1}{b^2+m^2}\,,\\
    \beta_m &= \frac{\beta b^2}{b^2+m^2}\,,\\
    \gamma_m &= \gamma \frac{\sqrt{2}b}{\pi}\frac{4m^3}{(4m^2-1)(b^2+m^2)}\,,\\
    \tilde\gamma_m &= \gamma \frac{\sqrt{2}b}{\pi} \frac{4m}{4m^2-1}\,,\\
    \delta_m &= \frac{64\sqrt{2}}{15\pi} \frac{b^2-m^2+1}{b^2+m^2}\,,\\
    \eta &= \frac{16\sqrt{2}}{5\pi}\,.
  \end{split}
\end{equation}
The original model is detailed in~\cite{charney-devore:1979}, and was
modified in~(\ref{eq:cdv}) to add additive Gaussian noise to each
degree of freedom. The model~(\ref{eq:cdv}) allows for two metastable
states, the so-called ``zonal'' state, and the ``blocked'' state,
alluding to the atmospheric blocking phenomena observed in
meteorology.

Application of the action minimization algorithm introduced in
section~\ref{sec:simpl-geom-minim} allows us to compute the most
likely transition trajectories in the small noise limit, $\eps\to0$,
and deduce the relative stability of the states. The results are shown
in figure~\ref{fig:cdv-transition}: Starting from the zonal state in
the upper left corner, snapshots of the streamfunction $\psi(x,y)$ are
shown along the transition trajectory in lexicographic order, arriving
at the blocked state. For comparison, the right collection of plots in
figure~\ref{fig:cdv-transition} shows the corresponding backward
transition from blocked to zonal state. Note that the backward
transition is not merely the time-reversed forward transition,
implying (as expected) a breaking of time-reversal symmetry and thus
demonstrating the non-equilibrium nature of the transition. The
relative stability of the two configurations can be quantified via
figure~\ref{fig:cdv-action}: The action to transition towards the
blocked state is far larger than the action to transition towards the
zonal state, meaning that the zonal state is exponentially preferred
in the low-noise limit.

\subsection{Instantons for stochastic partial differential equations}
\label{sec:inst-stoch-part}

Many systems of interest in physical applications have continuous
spatial variables, i.e.~do not fit the framework of
equation~(\ref{eq:1}). Instead, they are stochastic \emph{partial}
differential equations (SPDEs).  Applying the algorithm of
section~\ref{sec:simpl-geom-minim} to stochastic processes in
infinite-dimensional spaces is nevertheless largely done in
practice. The mathematical foundation is less clear in this case,
though, and a few comments are in order.

A stochastic partial differential equation, even in the simplest case
of additive Gaussian noise, is possibly ill-posed. Consider for example
\begin{equation} \label{eq:basic_spde}
  \partial_t U = B(U) + \sqrt{\epsilon}\, \eta(x,t)\,,
\end{equation}
where $U:[0,T]\times\RR^d \to \RR^m$ and $\eta$ denotes temporal white
noise. If the noise is also not smooth in space, for example if it is
white-in-space as well, $\EE \eta(x,t)\eta(x',t') =
\delta(t-t')\delta(x-x;)$, it is a non-trivial undertaking to make
sense of possible non-linear terms in the drift, $B(U)$, especially if
the spatial dimension is higher than one. Recent mathematical
breakthroughs~\cite{hairer:2014} specify a rigorous renormalization
procedure in specific cases. In regards to LDT, the main concern is
whether this renormalization procedure subsists in the limit
$\eps\to0$. For example, in~\cite{hairer-weber:2015}, it was discussed
for the stochastic Allen-Cahn equation (e.g.~$B(U) = U-U^3 + \kappa
\partial_x^2 U$) in 2 or 3 spatial dimensions, that indeed the rate
function corresponds to the naively assumed one,
\begin{equation}
  \label{eq:action-stochastic-AC}
  S_T(\phi) = 
  \begin{cases}
    \int_0^T \|\partial_t \phi - B(\phi)\|_{L^2}^2\,dt\,, & \text{if the integral converges,}\\
    \infty & \text{otherwise,}
  \end{cases}
\end{equation}
where $\|\cdot\|_{L^2}$ denotes the $L^2$-norm in the spatial
components. In the following, we will consider SPDE examples, but no
longer dwell upon the mathematical intricacies, instead assuming
that~(\ref{eq:action-stochastic-AC}) is valid.

If the rate function takes the form~(\ref{eq:action-stochastic-AC}),
all arguments put forward in the finite dimensional case can be
transferred to the SPDE situation, and a corresponding algorithm can
be constructed. In particular, a gradient descent of the form
introduced in section~\ref{sec:rare-event-algor} is still feasible,
with gradients of vectors replaced by functional derivatives of the
corresponding operators. Therefore,
equation~(\ref{eq:sgmam-dk-update}) to compute the descent direction
for the SPDE~(\ref{eq:basic_spde}) becomes
\begin{equation}
  \label{eq:sgmam-SPDE-dk-update}
  d^k = {\theta^k}' + \left(D_\phi B(\phi)\right)^\top \theta^k\,,
\end{equation}
where $A^\top$ is the $L^2$-adjoint of the (differential) operator $A$
and $D_\phi$ the functional derivative. Consider for example Burgers
equation with periodic boundary conditions, where
\begin{equation}
  B(U) = \nu \partial_x^2 U - U \partial_x U\,.
\end{equation}
Then $D_\phi B(\phi)$ is the operator
\begin{equation}
  D_\phi B(\phi) = \nu \partial_x^2 - (\partial_x U) - U \partial_x
\end{equation}
such that
\begin{equation}
  \left(D_\phi B(\phi)\right)^\top = \nu \partial_x^2 + U \partial_x\,.
\end{equation}
Recall that we can still compute $\theta^*$, i.e.~the minimizer of the
inner constrained optimization problem~(\ref{eq:e-problem}) via
\begin{equation}
  \theta^*(\phi) = \mu^{-1}\phi' - B(\phi) = \mu^{-1}\phi' - \nu \partial_x^2 \phi + \phi \partial_x \phi\,,
\end{equation}
where the last equality holds for the Burgers example with
spatio-temporal white noise.

In practice, equation~(\ref{eq:sgmam-SPDE-dk-update}) has to be
rewritten in order to be practical for the SPDE case, because the
involved high spatial derivatives come with stability conditions (CFL
conditions) that limit the time-step of the descent, and therefore the
convergence rate of the scheme. A detailed discussion of tricks and
optimizations for the SPDE case is given in~\cite{grafke-schaefer-vanden-eijnden:2017}.

\subsection{Example: The stochastic Burgers-Huxley equation}
\label{sec:exampl-stoch-burg}

\begin{figure}
  \begin{center}
    \includegraphics[width=246pt]{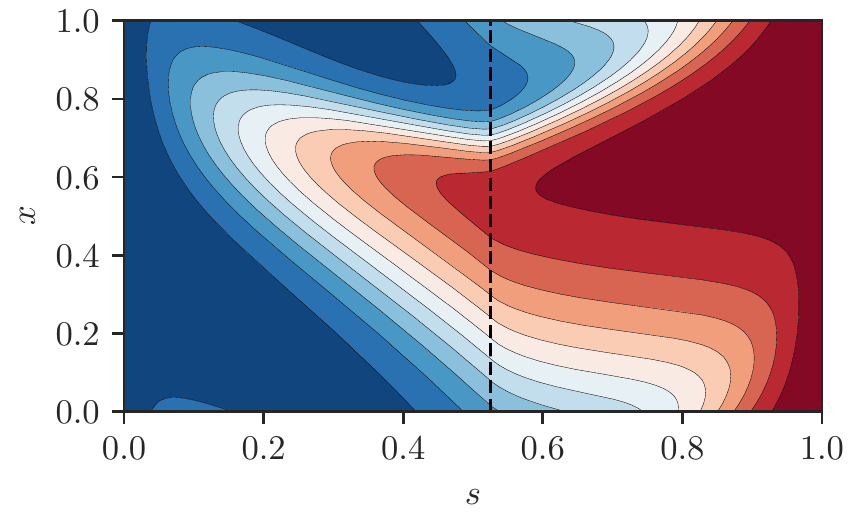}
  \end{center}
  \caption{Maximum likelihood transition pathway of the bi-stable
    Burgers-Huxley model, transitioning from $u=-1$ to $u=1$. The
    transition happens as Allen-Cahn like nucleation, but the critical
    nucleus forms as steepening, asymmetric shock-wave. The
    saddle-point, denoting the critical nucleus of the transition, is
    marked by a dashed line.\label{fig:burgers-huxley}}
\end{figure}

As example for a nonlinear SPDE, we consider the stochastic
Burgers-Huxley model,
\begin{equation}
  \partial_t u + \alpha u \partial_x u - \kappa \partial_x^2 u = f(u)
  + \sqrt{\eps} \eta(x,t)\,, x\in[0,1]\,,
\end{equation}
where $\alpha>0$ determines the strength of the nonlinear advection
term, $\kappa>0$ is the diffusion constant, and the boundary
conditions are periodic. The field $\eta(x,t)$ is spatio-temporal
white noise. For $f(u)=0$, this equation is the stochastic Burgers
equation, arising in compressible gas dynamics, traffic flows, and as
test-bed for turbulence. With the inclusion of a double-well reaction
term $f(u)=u-u^3$, the equation becomes metastable, with two spatially
homogeneous stable fixed points at $u=-1$ and $u=1$. The spatially
homogeneous solution $u=0$ is a fixed point as well, but depending on
the size of $\kappa$ might not be a saddle point with a single
unstable direction. Instead, for small enough~$\kappa$, we expect
Allen-Cahn like nucleation dynamics, but the nucleation must happen as
a Burgers-like steepening shock wave. Indeed, as
figure~\ref{fig:burgers-huxley} shows, this intuition is confirmed by
the numerics: The creation of the nucleus happens in a spatially
asymmetric way, and the nucleating seed travels in space. The spatial
resolution for the numerical computation is $N_x=256$, while the
temporal resolution is $N_s=100$.

\section{Rare event algorithms for expectations and extreme events}
\label{sec:rare-event-algor-1}

In section~\ref{sec:rare-event-algor} we discussed how to compute
noise-induced transition trajectories in bi-stable systems and thereby
estimate the rate of transitions between the two metastable states and
their relative likelihood. We chose to implement a global minimization
procedure based on the Maupertuis principle form of the action
functional to use the independence of the choice of parametrization to
our numerical advantage. Nevertheless, because both the initial
\emph{and} the final conditions of the transition trajectory are fixed, we
were unable to harness the Hamilton's equations of motion directly
(these would have to be solved by shooting methods, which are
inefficient or even ill-posed).

In this section, instead, we will concentrate on situations where it
is indeed feasible to solve the minimization problem by integrating
the coupled pair of equations of motion, or instanton equations,
to obtain the large deviation minimizer. As we will see, if
applicable, this approach comes with a couple of advantages of both
theoretical and numerical nature. In this section we will therefore
first review the class of algorithms based on solving Hamilton's
equations in section~\ref{sec:inst-expect}, that can be used to
compute instantons for expectations dominated by extreme
events. Examples are shown in section~\ref{sec:exampl-lotka-volterra}
applying this algorithm to a system with multiplicative Gaussian
noise, and furthermore in section~\ref{sec:exampl-extr-KdV}
demonstrating the use in an infinite dimensional system, with the
additional complication of degenerate forcing (i.e.~non-invertible
noise covariance matrix). We discuss connections to other fields in
section~\ref{sec:impr-conn-optim} and numerical details in
section~\ref{sec:impr-impl-cons}. A geometric variant of the numerical
scheme is introduced in section~\ref{sec:geom-vers-hamilt}, and
implemented for an example case in
section~\ref{sec:exampl-extr-grad-burgers}.

\subsection{Instantons for expectations and extreme events}
\label{sec:inst-expect}

For the stochastic process $X_t^\eps$ of equation~(\ref{eq:1}),
\begin{equation*}
  dX_t^\eps = b(X_t^\eps)\,dt + \sqrt{\eps} \sigma\,dW_t\,,
\end{equation*}
consider the random variable $F(X_T^\eps)$, where $F:\RR^d\to\RR$.
This random variable, also termed the ``observable'', acts only on the
final configuration of the process. We are interested in estimating
the tail scaling of its probability density, i.e.~in quantifying the
likelihood of extreme values of the observable. For example, assume
that $X_t^\eps$ is a stochastic model describing the interaction of
predator and prey in a habitat
(cf.~section~\ref{sec:exampl-lotka-volterra}). We set out to find the
probability of observing an abundance of prey. We might additionally
be interested in the most likely amount of predators at this unusual
configuration, and the historic development into this event. Or we
have a stochastic description of waves
(cf.~section~\ref{sec:exampl-extr-KdV}), and are interested in the
probability of observing high amplitude waves. Additionally we might
ask for the most likely shape of the wave at the moment of extreme
elevation, or possibly identify the evolution into the extreme wave
event to analyze it for possible mechanisms leading to the
amplification.

From the discussion in section~\ref{sec:introduction} we understand
that in the limit $\eps\to0$, the probability of observing the event
$F(X_T^\eps)=z$, subject to $X_0^\eps=x$, fulfills
\begin{equation}
  \label{eq:optimization-for-F}
  P(z) \asymp \exp(-\eps^{-1} \inf_{\phi\in\mathscr C_z} S_T(\phi))\,,
\end{equation}
where $\mathscr C_z = \{ \phi \in C([0,T],\RR^d)\ |\ \phi(0)=x,
F(\phi(T))=z\}$, i.e.~the set of continuous trajectories starting and
$x$ that observe the event. Let
\begin{equation}
  \label{eq:I}
  I(z) = \inf_{\phi\in\mathscr C_z} S_T(z)\,,
\end{equation}
and define
\begin{equation}
  \label{eq:Istar}
  I^*(\lambda) = \inf_{\phi\in\mathscr C} (S_T(\phi) - \lambda F(\phi(T)))\,,
\end{equation}
with $\mathscr C = \{\phi\in C([0,T],\RR^d)\ |\ \phi(0)=x\}$. Here,
minimization is not constrained at the final point, i.e.~$\mathscr C$
describes the set of continuous trajectories starting at $x$
regardless of their final point. The G\"artner-Ellis theorem states,
that $I^*(\lambda)$ and $I(z)$ are Fenchel-Legendre duals. This can be
motivated by rewriting
\begin{align*}
  I^*(\lambda) &= \inf_{\phi\in\mathscr C}(S_T(\phi) - \lambda F(\phi(T)))\\
  &= \inf_{z\in\RR}\inf_{\phi\in\mathscr C_z}(S_T(\phi)-\lambda F(\phi(T)))\\
  &= \inf_{z\in\RR}(\inf_{\phi\in\mathscr C_z} S_T(\phi) - \lambda z)\\
  &= \inf_{z\in\RR}(I(z) - \lambda z)\,.
\end{align*}
Effectively, the connection between $I(z)$ and $I^*(\lambda)$ allows
us to solve the minimization problem~(\ref{eq:Istar}) instead of the
original problem~(\ref{eq:I}). In terms of Hamilton's principle, the
variations of the argument of the infimum in equation~(\ref{eq:Istar})
with respect to $\phi$ gets one additional term that only applies at
the final point, so that the Hamilton's equations become
\begin{equation}
  \label{eq:eqmo-expect}
  \begin{cases}
    \dot\phi = \nabla_\theta H(\phi,\theta) & \phi(0)=x\\
    \dot\theta = -\nabla_\phi H(\phi,\theta)& \theta(T)=-\lambda \nabla F(\phi(T))\,.
  \end{cases}
\end{equation}
The difference with Hamilton's equations of the problem discussed in
section~\ref{sec:rare-event-algor}
\begin{equation}
  \label{eq:eqmo-classic}
  \begin{cases}
    \dot\phi = \nabla_\theta H(\phi,\theta) & \phi(0)=x,\ F(\phi(T))=z\\
    \dot\theta = -\nabla_\phi H(\phi,\theta)& \text{(no boundary conditions)}
  \end{cases}
\end{equation}
appears minuscule, but is profound: The $\phi$-equation
in~(\ref{eq:eqmo-classic}) has to be solved with initial \emph{and}
final condition, and therefore necessitates shooting methods which are
inefficient in high dimension (hence the alternative approach we took
in section~\ref{sec:rare-event-algor}). For~(\ref{eq:eqmo-expect}), on
the other hand, the equations for both $\phi$ and $\theta$ have
exactly one boundary condition each. It is natural to integrate the
$\phi$-equation forward in time, starting at $x$, while integrating
the $\theta$-equation backward in time, starting at
$-\lambda \nabla F(\phi(T))$. This direction of integration is the
only sensible one in the first place: Due to the conjugate momentum
equation containing the term $-(\nabla b(\phi))^\top$, a numerical
integration forward in time would be numerically unstable or even
ill-posed. An algorithm to find the instanton in this case then
consists of the following steps: Starting from the $k$-th guess
$\phi^k(t)$ for the instanton trajectory,
\begin{enumerate}[(i)]
\item solve the equation
  \begin{equation}
    \dot\theta = -\nabla_\phi H(\phi^k,\theta),\quad\theta(T)=-\lambda F(\phi^k(T))
  \end{equation}
  backward in time,
\item solve the equation
  \begin{equation}
    \dot\phi = \nabla_\theta(\phi,\theta),\quad\phi(0)=x
  \end{equation}
  forward in time to obtain the next guess $\phi^{k+1}$,
\item iterate until convergence.
\end{enumerate}
The convergence properties, stability and possible improvements of
this algorithm are discussed in section~\ref{sec:impr-conn-optim}.
Considering the dual problem~(\ref{eq:Istar}) instead of the original
one~(\ref{eq:I}) comes at a price: Instead of choosing directly the
value $z$ of the observable, instead we prescribe its dual $\lambda$,
and obtain the corresponding value of $z$ \emph{a posteriori}. In
other words, we loose the capability of computing the instanton for a
specific observable $z$. In practice, this is usually not a problem,
even though the map $z(\lambda)$ is not available in general:
Typically one is interested in the complete distribution of $P(z)$,
and therefore producing instantons for a whole range of $\lambda$
similarly covers a whole range of $z$. Alternatively, a
self-correcting version of the algorithm is easily implemented, where
$\lambda$ is adjusted on-the-fly to achieve the desired outcome $z$.

Note that $I^*(\lambda)$ is nothing but the limit of the scaled
cumulant generating function of the random variable $F(X_T^\eps)$,
i.e.
\begin{equation}
  I^*(\lambda) = \lim_{\eps\to0} \eps \log \EE \exp(\lambda F(X_T^\eps))\,.
\end{equation}
In this interpretation, we could call the instanton
solving~(\ref{eq:eqmo-expect}) also the instanton corresponding to the
expectation
\begin{equation}
  \EE \exp(\eps^{-1} \lambda F(X_T^\eps))
\end{equation}
in the limit $\eps\to0$. It is similarly possible to define
observables not only on the final point of the trajectory, but for
example of the form
\begin{equation}
  F(\{X_t^\eps\}) = \int_0^T \!\!f(X_t^\eps)\,dt\ \ \text{or}\ \  F(\{X_t^\eps\}) = \int_0^T \!\!\langle g(X_t^\eps), dW\rangle
\end{equation}
and perform similar arguments, leading to additional drift terms in
the conjugate momentum equation.

Finally, while the above arguments rigorously hold under suitable
conditions in the limit $\eps\to0$, it is common to loosen conditions
on the stochastic process and consider the case $\eps$ fixed, but
$\lambda\to\infty$. The intuition is that for large $\lambda$ only
extreme events of the process are considered, and a large deviation
principle might hold for the observable even for finite noise. One can
then write down an \emph{a priori} large deviation principle for the
random variable $F(X_T^\eps)$ and compute the instanton for large
values of $\lambda$ to probe the tail of the probability to observe
the event. It is in this sense that this approach can be considered as
instantons for \emph{extreme events}. They are commonly used in
practice, for example in fluid dynamics, where an equivalent algorithm
has been introduced by Chernykh and
Stepanov~\cite{chernykh-stepanov:2001}, which was applied to compute
instantons for Burgers, Navier-Stokes, and KPZ
equations~\cite{grafke-grauer-schaefer:2013,
  grafke-grauer-schaefer-etal:2015, grafke-grauer-schaefer:2015,
  zarfaty-meerson:2016}.

\subsection{Example: Extreme concentration of prey in the Lotka-Volterra model}
\label{sec:exampl-lotka-volterra}

\begin{figure}
  \begin{center}
    \includegraphics[width=246pt]{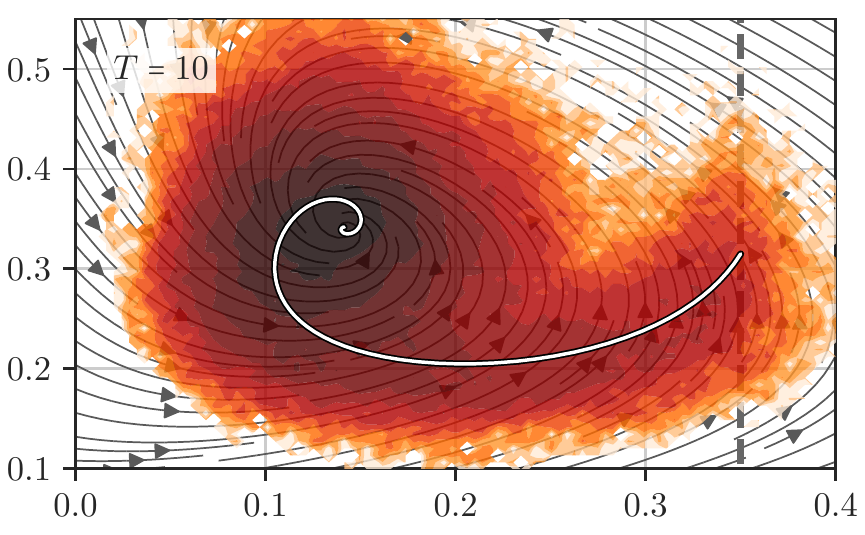}\\
    \includegraphics[width=246pt]{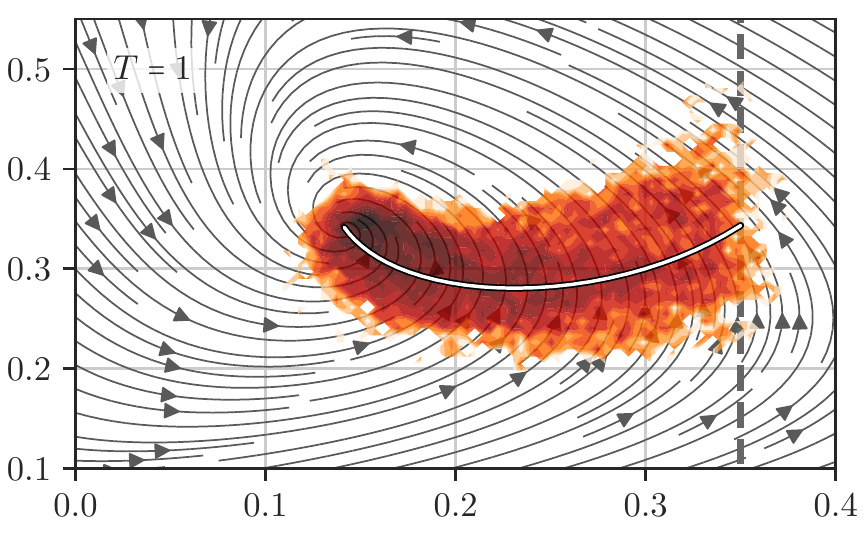}
  \end{center}
  \caption{Occurrence of extreme concentration of prey in the
    Lotka-Volterra model. The streamlines are showing the deterministic
    flow field. The heat-map shows the logarithm of a histogram of all
    trajectories starting at the fixed point that reach $a(T)=0.35$
    (regardless of $b(T)$). The white line depicts the instanton for
    the expectation $\EE \exp(-\lambda a(T))$. Even for finite $\eps$,
    the sample trajectories clearly cluster around the
    instanton. Shown are two different event times, $T=10$ (top) and
    $T=1$ (bottom).}
  \label{fig:lv}
\end{figure}

The Lotka-Volterra system, or predator-prey system, is frequently
used in biology as the simplest description of the interaction of two
species, one of which preys on the other. In its typical form, it is
considered without any fluctuations, but as it can be understood as
continuous limit of a network of reactions, it is clear how a noise
term can be derived as chemical Langevin equation.

To this end, consider a habitat with two species, the prey $A$ and the
predator $B$, where $A,B\in\ZZ^+$ denotes the number of individuals of
the respective species. We want to consider interactions between
individuals, modeled by the stoichiometric reaction network
\begin{subequations}
  \label{eq:lv-network}
  \begin{align}
    A &\xrightarrow{\alpha} A+A &&\text{(reproduction of prey)}\\
    A+B &\xrightarrow{\beta} B+B&&\text{(predation)}\\
    B &\xrightarrow{\gamma} \emptyset&&\text{(death of predator)}\\
    \emptyset &\xrightarrow{\delta} A&&\text{(migration of prey)}\\
    \emptyset &\xrightarrow{\delta} B&&\text{(migration of predator)\,.}
  \end{align}
\end{subequations}
Each of these is to be understood as a Poisson process with rates
$\alpha$ to $\delta$. The first three are standard in Lotka-Volterra,
the last two are added to model migration of both species from
neighboring habitats towards the considered location. This prevents
degeneracy of the forcing at extinct population levels and the
difficulty of absorbing boundary conditions.

Under the assumption that the typical populations $N$ are sufficiently
large, $\eps=N^{-1}\to0$, the chemical Langevin equation
corresponding to the reaction network~(\ref{eq:lv-network}) is
\begin{equation}
  \label{eq:SDE-trunc}
  \begin{cases}
    da = (-\beta ab + \alpha a + \delta)\,dt + \sqrt{\eps}\sqrt{\beta ab + \alpha a + \delta}\, dW_a\,,\\
    db = (\beta ab - \gamma b + \delta)\,dt + \sqrt{\eps}\sqrt{\beta ab + \gamma b + \delta}\, dW_b\,,
  \end{cases}
\end{equation}
where $a,b$ are functions from $[0,T]$ into $\RR^+$, denoting the
concentration of predator and prey in the habitat. The stochastic
fluctuations are white-in-time Gaussian and zero mean, but notably
multiplicative. Note that while this noise term indeed is consistent
with a central limit theorem for $N\to\infty$, it is actually not true
that this approximation is valid for large deviations as well. In
general, the LDP is sensitive to the non-Gaussian nature of the
stochastic process defined in~(\ref{eq:lv-network}), which has Poisson
statistics. We explain how to treat such non-Gaussian systems
correctly in section~\ref{sec:gener-non-gauss}, while here, for
simplicity, we are considering the multiplicative Gaussian
SDE~(\ref{eq:SDE-trunc}) as given.

We choose the interaction rates $\alpha, \beta, \gamma,$ and $\delta$
in a way that there exists a unique fixed point $(\bar a, \bar b)$ at
which concentrations of predators and prey are in
equilibrium. Concretely, we take $\alpha = 1$, $\beta = 5$, $\gamma =
1$, $\delta = 0.1$. Changing these parameters can produce more
complicated attractors, such as limit cycles, which we will not
investigate here. Instead, we are interested in the question of how
unlikely high concentrations of prey develop on different time frames
$T$ when the system starts at the fixed point $(\bar a, \bar b)$. For
that reason, we choose $F(a,b) = a(T)$, i.e.~condition on high values
of $a(T)$, regardless of $b(T)$.

Since this is the first time we encounter
multiplicative noise, a few comments are in order. For a system of the
form
\begin{equation}
  dX_t^\eps = b(X_t^\eps)\,dt + \sqrt{\eps}\sigma(X_t^\eps)\,dW_t\,,
\end{equation}
with $a(x)=(\sigma\sigma^\top)(x)$, the Hamiltonian is
\begin{equation}
  H(\phi,\theta) = \langle b(\phi),\theta\rangle + \tfrac12\langle \theta, a(\phi)\theta\rangle\,,
\end{equation}
so that an additional term enters the equation for the conjugate momentum,
\begin{equation}
  \dot \theta = -\nabla_\phi H(\phi,\theta) =-(\nabla b(\phi))^\top \theta + \langle \theta, \nabla_\phi a(\phi) \theta\rangle\,,
\end{equation}
where the last term is to be understood as $(\langle \theta,
\nabla_\phi a(\phi) \theta\rangle)_i = \sum_{j,k} \theta_j
\nabla_{\phi_i} a_{jk} \theta_k$. Consequently, the instanton
equations for the (stochastic) Lotka-Volterra model are
\begin{equation}
  \begin{cases}
    \dot a = -\beta ab + \alpha a + \delta + (\beta a b + \alpha a + \delta) p_a\\
    \dot b = \beta ab - \gamma b + \delta + (\beta a b + \gamma b + \delta) p_b\\
    \dot \theta_a = -(\alpha-\beta b)\theta_a -\beta b \theta_b + \tfrac12((\alpha+\beta b)\theta_a^2 + \beta b \theta_b^2)\\
    \dot \theta_b = \beta a \theta_a - (-\gamma+\beta a)\theta_b + \tfrac12(\beta a \theta_a^2 + (\gamma + \beta a)\theta_b^2)\,,\\
  \end{cases}
\end{equation}
which have to be solved with the boundary conditions $(a(0), b(0)) =
(\bar a, \bar b)$ and $(\theta_a(T), \theta_b(T)) = -\lambda \nabla F(a,b) =
(-\lambda, 0)$.

Figure~\ref{fig:lv} shows the result of applying the algorithm of
section~\ref{sec:inst-expect} to this system, and comparing to
Monte-Carlo sampling. Here, two different transition times are chosen,
$T=1$ and $T=10$. For $T=10$, the system has enough time to explore
around the fixed point, but it is obvious that the last portion of the
excursion, before it hits $a(T)=0.35$, clusters around the instanton
trajectory. In particular, the $b$-coordinate at which $a=0.35$ is
attained seems to be predicted reasonably well. For $T=1$, instead,
the transition trajectory needs to follow a different route, and the
endpoint $a=0.35$ will most likely be attained at higher concentration
of predators. Some points, such as $(a,b)=(0.25,0.3)$, are almost
never visited for $T=10$, but are very likely under $T=1$, which is
correctly predicted by the instanton computation. Note that the
heat-map depicting the empiric probability density in
figure~\ref{fig:lv} has a logarithmic color-map for the tails to
remain visible. Deviations from the optimal path are therefore very
unlikely indeed.

Parameters for $T=1$ are $\Delta t=10^{-2}$, $\eps=0.005$ and
$\lambda=0.4209$, and for $T=10$ are $\Delta t=10^{-2}$, $\eps=0.004$,
and $\lambda=0.2106$. Roughly $5\cdot 10^{6}$ trajectories were
sampled for the Monte-Carlo estimate.

\subsection{Example: Extreme amplitudes in the Korteweg-de Vries equation}
\label{sec:exampl-extr-KdV}


\begin{figure}
  \includegraphics[width=246pt]{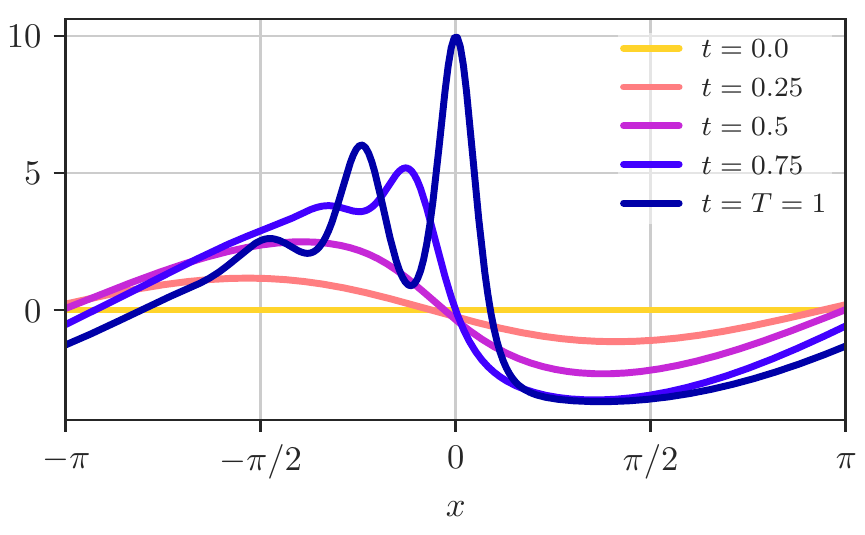}
  \caption{Evolution of the Korteweg-de Vries instanton into a large
    amplitude at the final time $T=1$, starting from rest and forcing
    only the largest Fourier mode of the system.}
  \label{fig:kdv}
\end{figure}

Consider for the field $u(x,t):[0,2\pi]\times[0,T]\to\RR$ the stochastic
partial differential equation
\begin{equation}
  \partial_t u + u\partial_x u + \kappa \partial_x^3 u - \nu \partial_x^2 u = \eta(x,t)\,,\quad u(x,t\!=\!0)=0\,,
\end{equation}
with periodic boundary conditions in space, and $x\in[0,2\pi]$. This
is a modification of the standard Korteweg-de Vries equation that
describes the evolution of shallow water surface waves. To this, we
added energy input through the forcing $\eta$ and energy dissipation
through a diffusion term with viscosity $\nu$. For the forcing, we
demand that, in Fourier space,
\begin{equation}
  \EE \hat\eta_k(t) \hat\eta_q(t') = \delta(t-t')\hat\chi_{q-k}\,,
\end{equation}
where $\hat\chi:\ZZ\to\RR$ is the forcing spectrum and $\hat\eta_k$ is
the $k$-th mode of the Fourier transform of $\eta$.

Intuitively, for a $\hat\chi_k$ with compact support only for small $k$,
the forcing $\eta(x,t)$ inserts energy on large scales, and the
nonlinearity transfers those to smaller scales, on which dispersion
and dissipation act on them. We are interested how this nonlinear
cascading effect produces waves of extreme amplitude. To this effect,
we choose an observable
\begin{equation}
  \label{eq:observable-KdV}
  F(u(x,T)) = (\phi_\Delta \star u)(x,T)\,,
\end{equation}
with $\phi_\Delta(x) = A \exp\left(-x^2/\Delta^2\right)$, and
$\star$ denoting spatial convolution. For small $\Delta$, this
observable selects high amplitudes in close proximity to $x=0$,
i.e.~at the center of the domain, and therefore generates high wave
elevations at this position. As forcing spectrum, we want
\begin{equation}
  \chi_k=
  \begin{cases}
    1&\text{if } |k|=1\\
    0&\text{otherwise}\,,
  \end{cases}
\end{equation}
which inserts energy only into the largest mode of the system. This is
the first time we consider \emph{degenerate} forcing, in that only a
subset of the available degrees of freedom are forced, or,
equivalently, the noise covariance matrix $a$ of~(\ref{eq:1}) is not
invertible. This poses practical problems for algorithms based on
global minimization discussed in section~\ref{sec:rare-event-algor},
where heavy use is made of either the $a$-norm, or $\theta$ is
expressed as $\theta=a^{-1}(\dot\phi-b(\phi))$. For these algorithms,
the degenerate forcing introduces additional stiff constraints for the
unforced modes, as those effectively behave deterministically (and
thus are attained with infinite action if they deviate from the
deterministic behavior). For the Hamilton's equations, and the
algorithm discussed in this section, the noise correlation is never
inverted, and degenerate forcing can be treated without extra effort.

The instanton equations corresponding to the posed problem are
\begin{equation}
  \label{eq:KdV-instanton-equations}
  \begin{cases}
    \partial_t u + u\partial_x u + \kappa \partial_x^3 u - \nu \partial_x^2 u = \chi\star\theta,\ \  u(x,t\!=\!0)=0\,,\\
    \partial_t \theta + u\partial_x \theta + \kappa \partial_x^3 \theta + \nu \partial_x^2 \theta = 0,\ \ \theta (x,t\!=\!T)=-\lambda \phi_\Delta(x)\,,
  \end{cases}
\end{equation}
where $\chi(x)$ is the inverse Fourier transform of the forcing
spectrum. The instanton computed by solving
equations~(\ref{eq:KdV-instanton-equations}) is depicted in
figure~\ref{fig:kdv}. It is clearly visible that a high final
amplitude around $x=0$ is achieved by a combination of non-linear
advection and dispersion. Additionally, the final configuration
clearly contains Fourier modes different from $|k|=1$, implying that
indeed the non-linearity cascaded energy into higher modes in a way to
optimize the final amplitude. Note also that because of $\Delta\ll1$,
we are merely demanding a large amplitude at $x=0$, but leave the
rest of the wave form unconstrained. The elevation profile in the rest
of the domain is chosen in a most likely manner, and the curves shown
in figure~\ref{fig:kdv} can be interpreted as the prototypical way of
forming the considered amplitude. The model parameters are $T=1$,
$\Delta=10^{-1}$, $\alpha=\kappa=4\cdot10^{-2}$, $\lambda=1$, and
$A=0.25$. The numerical parameters are $N_x=256$, $N_t=1000$, and
$\Delta t = 10^{-3}$.

\subsection{Connections to optimal control}
\label{sec:impr-conn-optim}

It is instructive to formulate the optimization
problem~(\ref{eq:optimization-for-F}) in the language of optimal
control: We are interested in finding the \emph{optimal control}
$p:[0,T]\to\RR^d$ such that for $X\in\RR^d$, the system
\begin{equation}
  \label{eq:X-constraint}
  \dot X(t) = b(X(t)) + p(t), \quad X(0)=x\,,
\end{equation}
has the desired outcome, $F(X(T))=z$. We penalize large values of $p$
by choosing
\begin{equation}
  J(p) = \tfrac12\int_0^T |p(t)|^2\,dt
\end{equation}
as cost function. In other words, we are searching for the optimal
noise realization $p$ to drive the system into a final state where
$F(X(T))=z$. To obtain a minimization procedure that honors the
constraints given by the observable and
equation~(\ref{eq:X-constraint}), we introduce Lagrange multipliers
$\mu\in[0,T]\times\RR^d$ and $\lambda\in\RR$, such that we attempt to
minimize
\begin{align*}
  E(p) =&\ \frac12\int_0^T |p(t)|^2\,dt + \lambda F(X(T))\\& + \int_0^T \langle \mu, \dot X-b(X)-p\rangle\,dt + \mu(0)(X(0)-x)\,.
\end{align*}
Its total variation is given by
\begin{align*}
  \delta E(p) &= \langle p\!-\!\mu,\delta p\rangle + \langle\dot X\!-\!b(X)\!-\!p,\delta \mu\rangle + \langle X(0)\!-\!X,\delta \mu(0)\rangle\\
  &+ \lambda \langle \nabla F(X(T)), \delta X(T)\rangle + \langle -\dot\mu-(\nabla b(X))^\top \mu,\delta X\rangle\\
  &+ \langle \mu(T),\delta X(T)\rangle - \langle \mu(0),\delta X(0)\rangle\,.
\end{align*}
We can read of the desired conditions to fulfill the constraints as
\begin{equation}
  \label{eq:hamiltons-control}
  \begin{cases}
    \dot X= b(X) + p,& X(0)=x\\
    \dot\mu = -(b(X))^\top\mu,&\mu(T)=-\lambda\nabla F(X(T))\,,
  \end{cases}
\end{equation}
and the gradient of the cost functional $E(p)$ with respect to the
control $p$ is then given as
\begin{equation}
  \label{eq:dEdp}
  \frac{\delta E(p)}{\delta p} = p-\mu\,.
\end{equation}
We immediately identify that the conjugate momentum $\theta$ is the
variable $\mu$ in optimal control, often termed the \emph{adjoint}
variable. Second we realize that the forward and adjoint equations are
identical to the instanton equations. Therefore, the iterative
algorithm given in section~\ref{sec:inst-expect} is nothing but a
gradient descent for the cost functional $E(p)$, with step length
$1$. This not only answer questions about (local) convergence of the
algorithm of section~\ref{sec:simpl-geom-minim}, but furthermore
allows to improve stability and order of convergence of the
algorithm. First, it is almost always necessary to adjust the step
size for each iteration according to a line search strategy to achieve
convergence. Second, one might consider pre-conditioning, to allow for
faster convergence. Lastly, the computation of the descent direction
$-\delta E/\delta p$ from equation~(\ref{eq:dEdp}) allows to construct
higher order optimization algorithms, such as nonlinear conjugate
gradient or quasi-Newton methods.

Note that, similar to the argument above, for practical reasons we
choose to not consider variations with respect to $\lambda$, and
instead consider $\lambda\in\RR$ given \emph{a priori} to establish a
mapping $\lambda(z)$ from $z(\lambda) = F(X^*(T))$, where $X^*(T)$
depends on $\lambda$ through the boundary condition of the adjoint
equation~(\ref{eq:hamiltons-control}).

\subsection{Improvements and implementation considerations}
\label{sec:impr-impl-cons}

A few remarks are in order to point out possible improvements and
implementation concerns when solving Hamilton's equations.
\begin{itemize}
\item While solving a global minimization problem as introduced in
  section~\ref{sec:rare-event-algor} necessitates a complicated
  procedure to compute descent directions, the solution of Hamilton's
  equations put forward in this section usually comes at a much lower
  \emph{implementation} cost: Given a stochastic problem at hand, one
  likely has already available an efficient solver of the forward
  equation, just replacing stochastic noise with a function of the
  conjugate momentum. The backward equation (auxiliary equation,
  adjoint equation), on the other hand, is often available as well for
  professional software packages, usually from automatic
  differentiation, in order to quantify the uncertainty from the
  adjoint field $\mu$. In this case, a computation of the instanton
  might be achieved in a truly ``black-box'' form, where the iterative
  solution of the Hamilton's equations can be considered as a pure
  outer problem.
\item The mutual dependency of the forward and backward equations
  necessitates in principle that the whole trajectory is stored in its
  entirety. While this is usually feasible for finite-dimensional
  problems, it quickly becomes prohibitive in terms of memory
  requirement when talking about SPDEs in many dimensions, where
  usually the storage requirements are chosen to be of the order of
  magnitude of a single state of the field variables, and not a
  continuous trajectory. In optimal control, this restriction is
  usually overcome via \emph{checkpointing} mechanisms, where one does
  not save to memory a complete trajectory, but instead only retains
  checkpoints, from which subsequent states can be deduced. In its
  most efficient form, this checkpointing can be implemented in a
  recursive manner, leading generally to memory requirements of
  $O(\log N_t)$ instead of the naive $O(N_t)$, where $N_t$ is the
  number of discretization points in time. Some details of the
  application of this method to instanton computations is laid out
  in~\cite{grafke-grauer-schindel:2015}.
\item The nature of the conjugate momentum as adjoint variable, as
  discussed in section~\ref{sec:impr-conn-optim}, highlights its
  \emph{adjointness} to the main field variable. Would one discretize
  the action, and consider forward and backward equation as their
  discrete variation, this state of affairs would necessitate the
  usage of a special temporal integrator for the numerical solution of
  the adjoint equation, namely a temporal integrator that is adjoint
  to the forward integrator. Some time integrators have the property
  of being self-adjoint, and thus can be used for both equations. The
  failure to use a correct pair of integrators generally results in
  the failure to converge to the minimum of the cost function.
\end{itemize}

\subsection{Geometric version of the Hamilton's equations}
\label{sec:geom-vers-hamilt}

As discussed in section~\ref{sec:introduction}, many questions,
including the computation of the quasi-potential, necessitate a
minimization not only over all possible paths $\phi(t)$, but also over
all possible time intervals $T>0$---for example, this is needed to
calculate expectations with respect to the invariant measure of the
process, assuming it exists.  For MAM-type algorithms, the additional
complication the minimization over $T>0$ introduces is resolved by
invoking Maupertuis principle and focusing on the computation of the
\emph{geometric} minimizer, i.e.~realizing that the minimizing
trajectory can be computed without explicit reference to its
parametrization.

For algorithms based on the Hamilton's equation, similar ideas and
extensions exists~\cite{grafke-grauer-schaefer-etal:2014}: Instead of
solving the original Hamilton's equations~(\ref{eq:5}), we can choose
a reparametrization $s(t)$, and consider Hamilton's equations in this
parameter,
\begin{equation}
  \label{eq:geometric-hamiltonian}
  \begin{cases}
    \phi' = \mu \nabla_\theta H(\phi,\theta)\,\\
    \theta' = \mu \nabla_\phi H(\phi,\theta)\,,
  \end{cases}
\end{equation}
where the prime denotes derivatives with respect to $s$ and $\mu =
dt/ds$. Now, as pointed out in section~\ref{sec:simpl-geom-minim},
$\mu$ can be interpreted as Lagrange multiplier enforcing the
Hamiltonian constraint, and is available as
\begin{equation}
  \mu = \frac{\|\phi'\|^2_\sim}{\langle \phi',\nabla_\theta H\rangle_\sim} = \frac{\|\phi'\|_a}{\|b(\phi)\|_a}\,,
\end{equation}
where the much simpler second form only holds for Gaussian additive
noise of the form~(\ref{eq:1}) (compare
section~\ref{sec:simpl-geom-minim}).

\subsection{Example: Extreme gradients of the stochastic Burgers equation}
\label{sec:exampl-extr-grad-burgers}

\begin{figure}
  \includegraphics[width=246pt]{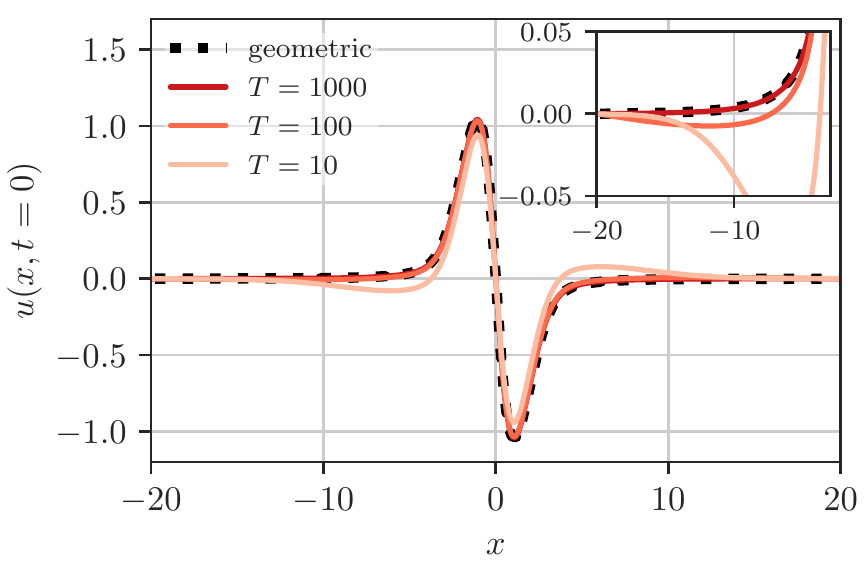}
  \caption{Comparison of the final condition of the instanton,
    $u(x,t=0)$, conditioning on extreme gradients in the origin, for
    the geometric parametrization and physical time parametrizations
    for varying $T$. As visible in the inset, only for $T=1000$,
    secondary extrema disappear.\label{fig:burgers_uends}}
\end{figure}
As an example, following~\cite{grafke-grauer-schaefer-etal:2014}, consider for the field
$u(x,t):[-L/2,L/2]\times[-T,0]\to\RR$ the stochastic Burger's equation,
\begin{equation}
  \label{eq:stochastic-burgers}
  \partial_t u + u \partial_x u - \nu \partial_x^2 u = \eta(x,t), \quad u(x,t=-T)=0\,,
\end{equation}
with periodic boundary conditions in space. Here, we consider a noise
term $\eta$ that is white in time, but has a finite correlation length
in space,
\begin{equation}
  \EE \eta(x,t)\eta(x',t') = \delta(t-t') \chi(x-x')\,,
\end{equation}
and we prescribe the specific correlation in Fourier space of
\begin{equation*}
  \hat\chi(k) = k^2 \exp(-k^2/2) \mathcal H(k_c - |k|)\,,
\end{equation*}
where $\mathcal H$ denotes the Heaviside step function. In effect, the
forcing correlation has the shape of a ``Mexican hat'' function, with
cut-off wave number $k_c$. Effectively,
equation~(\ref{eq:stochastic-burgers}) can be considered as a test-bed
for turbulence, where energy is inserted on large scales due to the
forcing, and then cascades to smaller scales via the nonlinearity,
where it dissipates. We focus on events that lead to a strong negative
gradient at the final time, $t=0$, and therefore choose an observable
\begin{equation}
  F(u(x,0)) = (\phi_\Delta \star \partial_x u)(x,0)\,,
\end{equation}
where, identically to the KdV-case in (\ref{eq:observable-KdV}),
$\phi_\Delta$ mollifies on scales $\Delta$, so that here we
concentrate on high gradients at the origin.

In order to probe for events on the invariant measure
of~(\ref{eq:stochastic-burgers}), we want to consider the limit
$T\to\infty$, and therefore need to either consider extremely large
time intervals $T$, or alternatively employ the geometric variant of
the Hamiltonian formalism as proposed in
(\ref{eq:geometric-hamiltonian}), where the norm is induced by the
covariance $\chi(x)$, i.e.~$\|v\| = \langle v, \chi^{-1}(x)
v\rangle^{1/2}$ on its support. For technical details,
see~\cite{grafke-grauer-schaefer-etal:2014}. Given this setup,
figure~\ref{fig:burgers_uends} compares the final condition of the
instanton between finite times $T$ and the limit $T\to\infty$ obtained
in the geometric formalism. It demonstrates how, for choices $T=10$ or
$T=100$, unphysical secondary maxima are present (compare inset of
figure~\ref{fig:burgers_uends}), that disappear in the infinite time
case, and with $T=1000$. Similarly, one can look at the value of the
Hamiltonian $H(u,p)$, which necessarily disappears for large $T$,
$\lim_{T\to\infty} H = 0$, and consider the quantity $\max(|H|)$ along
the instanton trajectory as measure of the numerical error of the
discretization. In this metric, the geometric variant for the same
number of discretization points in time, is roughly $10^4$ times more
accurate than the naive parametrization with physical
time~\cite{grafke-grauer-schaefer-etal:2014}.

\section{Generalizations to the non-Gaussian case}
\label{sec:gener-non-gauss}

In all above considerations and examples, we always considered the
presence of an LDP for an SDE, either with additive or with
multiplicative Gaussian noise. The class of stochastic systems
treatable with algorithms of the above form is far larger, though, as
is evidenced by the fact that both MAM-style global minimization of
section~\ref{sec:rare-event-algor} and the algorithms based on the
solution of Hamilton's equations of
section~\ref{sec:rare-event-algor-1} are written in terms of a generic
large deviation Hamiltonian. In this section, we therefore intend to
broaden the scope by demonstrating how more generic large deviation
Hamiltonians are obtained, and corresponding instantons can be
computed with the above algorithms. In particular, the Gaussianity of
the underlying stochastic process is reflected in the fact that the
Hamiltonian is \emph{quadratic} in its conjugate momentum. Other
processes, most notably those that result from limits of continuous
time Markov jump processes (MJPs), generally lead to a non-quadratic
Hamiltonian.

\subsection{Large deviation principles as WKB approximation}
\label{sec:WKB}

Consider a homogeneous continuous time Markov jump process $X_t$,
$t\in[0,T]$, with state space $\mathcal E$. The process is completely
characterized by its generator $\mathcal L$, which allows us to write
down its backward Kolmogorov equation (BKE) as
\begin{equation}
  \label{eq:BKE}
  \partial_t f + \mathcal Lf = 0\,, f(T)=\phi
\end{equation}
for
\begin{equation}
  f(T-t,n) = \EE^n \phi(X_t),
\end{equation}
where $n\in\mathcal E$, $f:[0,T]\times\mathcal E\to\RR$, and
$\phi:\mathcal E \to \RR$. For concreteness, consider as a state space
a counting space $\mathcal E = \ZZ_+^N$ for $N\in\NN$ different
species, where individuals of each species can interact via $R\in\NN$
independent Poisson processes, manipulating the number of individuals
of each species. For each of the $R$ different interactions, the
number of individuals changes from $n$ to $n+\nu_r$ with a rate
$\tilde a_r(n)$, $r\in\{1,\dots,R\}$. Then, the corresponding
generator reads
\begin{equation}
  (\mathcal L f)(n) = \sum_{r=1}^R \tilde a_r(n) \left(f(n+\nu_r)-f(n)\right)\,.
\end{equation}
Rescaling this into new variables $x=n/M$, where $M$ is a typical
number of individuals, we can expand in $\eps=M^{-1}$ to obtain a
large deviation principle in the limit of many individuals. To this
end, consider the generator in the rescaled variables,
\begin{equation}
  (\mathcal L^\eps f)(x) = \sum_{r=1}^R a_r(x) \left(f(x+\eps \nu_r) - f(x)\right)\,,
\end{equation}
where $a_r$ is defined on the rescaled variables. We now evaluate this
rescaled generator onto a function of the form $\exp(\eps^{-1} g(x))$
and rescale time with $\eps$ appropriately. This corresponds to a WKB
approximation of the BKE, or equivalently to the method of Feng and
Kurtz~\cite{feng-kurtz:2006}, and yields
\begin{equation}
  \partial_t g(x) + \sum_{r=1}^R a_r(x) \left( e^{\eps^{-1}(g(x+\eps \nu_r)-g(x))}\right)=0\,,
\end{equation}
which can be expanded, to leading order in $\eps$, into
\begin{equation}
  \label{eq:pre-HJ}
  \partial_t g(x) + \sum_{r=1}^R a_r(x) \left( e^{\langle \nu_r,\nabla g(x)\rangle}-1\right)=0\,.
\end{equation}
Equation~(\ref{eq:pre-HJ}) can be interpreted as Hamilton-Jacobi
equation
\begin{equation}
  \label{eq:HJ}
  \partial_t g(x) + H(x, \nabla g(x))=0\,,
\end{equation}
with
\begin{equation}
  \label{eq:HH}
  H(x,\theta) = \sum_{r=1}^R a_r(x) \left( e^{\langle \nu_r,\theta\rangle}-1\right)\,.
\end{equation}
The Hamiltonian~(\ref{eq:HH}) is precisely the large deviation
Hamiltonian in that the large deviation rate function is the time
integral of its Fenchel-Legendre transform. The
Hamiltonian~(\ref{eq:HH}) is furthermore a prime example of a
non-quadratic Hamiltonian.

Note that applying the same method to the generator of the
SDE~(\ref{eq:1}),
\begin{equation}
  (\mathcal L^\eps f)(x) = \langle b(x), \nabla\rangle f(x) + \tfrac\eps2 \sum_{i,j=1}^{d}a_{ij}\nabla_i\nabla_j f(x)
\end{equation}
recovers exactly the expected Hamiltonian~(\ref{eq:Hamiltonian-SDE})
for the leading order in $\eps$.

\subsection{Example: Genetic switch}
\label{sec:exampl-genet-switch}

\begin{figure}
  \begin{center}
    \includegraphics[width=248pt]{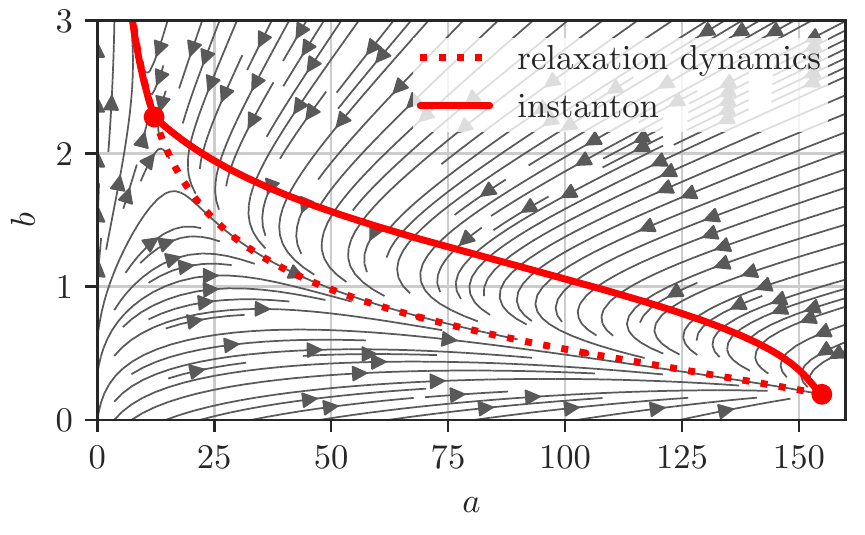}
  \end{center}
  \caption{Instanton for the non-Gaussian genetic switch. The arrows
    denote the direction of the deterministic flow, the red solid line
    depicts the minimizer, the red dashed line the relaxation path
    from the saddle. Red dots are located at the fixed points (stable
    and unstable). The whole figure is a zoom into the uphill region,
    the other stable fixed point is far up the upper left
    corner.\label{fig:genetic}}
\end{figure}

As an example for instantons of a non-Gaussian LDP for a continuous
time MJP of the form of section~\ref{sec:WKB}, we want to consider a
simplified model of a genetic switch: Inside a bacterium, plasmids
contain genes that encode two different proteins, $A$ and $B$. Each
protein is able to form polymers that inhibit the production of the
other protein, respectively. In the emerging situation, the cell may
exist in a state close to one of two fixed points: Either protein $A$
dominates, and inhibits the production of protein $B$, while the
production of $A$ remains high. Alternatively, protein $B$ dominates,
and inhibits the production of $A$. Rarely, fluctuations may arise
that push the system from one fixed point to the other.

We choose to model this system by describing the concentrations of
proteins $A$ and $B$ by $a\in\RR_+$ and $b\in\RR_+$,
respectively. There are four reactions in total, namely production and
degradation of $A$ and $B$, leading to the reaction network
\begin{subequations}
  \begin{align}
    \emptyset &\xrightarrow{p_A(b)} A && \text{(production of $A$)}\\
    \emptyset &\xrightarrow{p_B(a)} B && \text{(production of $B$)}\\
    A &\xrightarrow{d_A(a)} \emptyset && \text{(degradation of $A$)}\\
    B &\xrightarrow{d_B(b)} \emptyset && \text{(degradation of $B$)}
  \end{align}
\end{subequations}
with rates
\begin{equation}
  \begin{aligned}
    p_A(b) &= C/(1+b^3),& p_B(a) &= D/(1+a),\\
    d_A(a) &= a,& d_B(b)&=b\,.
  \end{aligned}
\end{equation}
The corresponding large deviation Hamiltonian, using
equation~(\ref{eq:pre-HJ}), is thus
\begin{equation}
  \label{eq:genetic-switch-H}
  \begin{aligned}
    H(a,b,\theta_a,\theta_b) &= \frac{C}{1+b^3}(e^{\theta_a}-1) +
    a(e^{-\theta_a}-1) \\&\ + \frac{D}{1+a}(e^{\theta_b}-1) +
    b(e^{-\theta_b}-1)\,.
  \end{aligned}
\end{equation}
The minimizer for this setup, as well as the relaxation paths from the
saddle, are depicted in figure \ref{fig:genetic}. They are computed by
implementing the algorithm presented in
section~\ref{sec:simpl-geom-minim} for the
Hamiltonian~(\ref{eq:genetic-switch-H}) for the transition between the
two fixed points. The model parameters here are chosen to be $C=156$
and $D=30$.

\section{Systems with random parameters and extreme events}
\label{sec:systems-with-random}

Up to now, all discussions in the previous sections concerned sample
path large deviations, where a stochastic process realizes a rare
event almost surely by following a trajectory that minimizes the
corresponding action functional. In this section, we are focusing on
a related, but different setup of a dynamical system with random
parameters. Given a distribution of the random parameters, we want to
reason about probabilities to observe certain events, and again
characterize the rare ones by dominating configurations of
parameters. To this effect, consider for $u:[0,T]\to\RR^d$ the
dynamical system
\begin{equation}
  \label{eq:dynsys}
  \partial_t u = b(u,\theta),\quad u(t=0)=u_0(\theta)\,,
\end{equation}
where $\theta\in\Omega\subseteq\RR^M$ is the set of $M$ random real
parameters, distributed according to a measure $\mu$. Since the
initial conditions and the drift of equation~(\ref{eq:dynsys}) depend
on the random parameters, the solution is a random variable, denoted
by $u(\cdot,\theta)$. We can then try to quantify the probability of
an observable exceeding a threshold $z\in\RR$, for example at the
final time, or integrated over time, or as temporal maximum, i.e.
\begin{equation}
  \label{eq:random-initial-F}
  P_T(z) = P(F(\theta)\ge z)\,, F(\theta) =
  \begin{cases}
    f(u(T,\theta))&\text{or}\\
    \int_0^T f(u(t,\theta))\,dt&\text{or}\\
    \displaystyle\max\limits_{0\le t\le T} f(u(t,\theta))\,.
  \end{cases}
\end{equation}

\subsection{Large deviations for systems with random parameters}
\label{sec:large-devi-syst}

Indeed, if in the limit of large $z$ this probability becomes small,
$\lim_{z\to\infty} P_T(z)=0$, then under some additional assumptions
on $F:\Omega\to\RR$ we have a large deviation principle of the form
\begin{equation}
  \label{eq:random-initial-LDP}
  P_T(z) \asymp \exp(-\min_{\theta\in\Omega(z)} I(\theta))\,,
\end{equation}
where the set $\Omega(z)\subseteq\Omega$ is the set of permissible
random parameters,
\begin{equation}
  \Omega(z) = \{\theta\in\Omega | F(\theta)\ge z\}\,,
\end{equation}
and the rate function $I(\theta)$ is obtained as the Legendre
transform of the cumulant generating function of $\theta$,
\begin{equation}
  I(\theta) = \max_{\eta} (\langle \eta,\theta\rangle - S(\eta))\,,
\end{equation}
for
\begin{equation}
  S(\eta) = \log \EE \exp\langle \eta,\theta\rangle = \log \int_\Omega \exp\langle \eta,\theta\rangle d\mu(\theta)\,.
\end{equation}
The minimizer of $I(\theta)$ in $\Omega(z)$, i.e.
\begin{equation}
  \label{eq:random-initial-instanton}
  \theta^*(z) = \argmin_{\theta\in\Omega(z)} I(\theta)\,,
\end{equation}
dominates the occurrence of the event, and is an \emph{instanton} in
this sense. Since we are considering only finite dimensional $\Omega$,
the corresponding optimization
problem~(\ref{eq:random-initial-instanton}) has to be solved in a
generally smaller search space. Equivalently, here, the instanton is
not a preferred trajectory of the system, but instead the maximum
likelihood set of parameters that lead to the event. Of course, given
any $\theta\in\Omega$, there is a unique trajectory $u(\theta)$
solving~(\ref{eq:dynsys}) associated to it, so that $u(\theta^*)$
represents the most likely trajectory of the system to realize the
rare event. The proof of the large deviation
principle~(\ref{eq:random-initial-LDP}) is carried out
in~\cite{dematteis-grafke-vandeneijnden:2018}.

From a numerical perspective, we can again solve the constrained
optimization problem~(\ref{eq:random-initial-instanton}) by instead
considering a Lagrange multiplier $\lambda\in\RR$. For example, for
the first of the three cases in~(\ref{eq:random-initial-F}), we
attempt to minimize the objective function
\begin{equation}
  \label{eq:objective-function-random-initial}
  E(u(T,\theta),\theta) = I(\theta) - \lambda f(u(T,\theta))\,.
\end{equation}
Via the Jacobian $J_{ij}(t) = \partial u_i(t)/\partial\theta_j$,
i.e.~the variation of the current configuration with respect to the
random parameters, one can express the gradient as
\begin{equation}
  \label{eq:nabla-E-in-J}
  \nabla_\theta E = \nabla_\theta I - \lambda J^\top(T,\theta)\partial_u f(u(T,\theta))\,,
\end{equation}
where the Jacobian is available through
\begin{equation}
  \label{eq:J}
  \partial_t J = (\partial_u b) J + \partial_\theta b,\quad J(0) = \partial_\theta u_0\,.
\end{equation}
While integrating the forward equation~(\ref{eq:dynsys}) and the
Jacobian~(\ref{eq:J}) allows us to evaluate the gradient for a given
$\theta$, note that $J:[0,T]\to\RR^{d\times d}$ is quite costly to
compute. Instead, we can again fall back to an adjoint formulation to
overcome this limitation. To this effect, consider the adjoint field
$\mu:[0,T]\to\RR^d$, subject to the adjoint equation
\begin{equation}
  \label{eq:random-initial-adjoint}
  \partial_t \mu = -(\partial_u b)^\top \mu,\ \mu(T) = \lambda \partial_u f(u(T,\theta))\,.
\end{equation}
Since, using equations~(\ref{eq:J})
and~(\ref{eq:random-initial-adjoint}), we have
\begin{equation*}
  \partial_t(J^\top \mu) = (\partial_\theta b)^\top \mu\,,
\end{equation*}
it follows that
\begin{equation*}
  \int_0^T \!\!(\partial_\theta b)^\top \!\mu\,dt = (J^\top\!\mu)\big|_0^T = \lambda J^\top\!(T) \partial_u f(u(T)) - (\partial_\theta u_0)^\top \!\mu(0)\,,
\end{equation*}
and thus, the gradient~(\ref{eq:nabla-E-in-J}) is computable without
referring to the Jacobian as
\begin{equation}
  \label{eq:nabla-E-in-mu}
  \nabla_\theta E = \nabla_\theta I - (\partial_\theta u_0)^\top \mu(0) - \int_0^T (\partial_\theta b)^\top \mu \,dt\,.
\end{equation}
In total, the gradient of the objective
function~(\ref{eq:objective-function-random-initial}) can be computed at a given $\theta$ in three steps:
\begin{enumerate}[(i)]
\item Integrate the forward equation,
  \begin{equation*}
    \partial_u = b(u,\theta),\quad u(0)=u_0(\theta)\,,
  \end{equation*}
\item Compute the adjoint field $\mu$ by integrating
  \begin{equation*}
    \partial_t \mu = -(\partial_u b)^\top \mu, \quad\mu(T) = \lambda \partial_u f(u(T))\,,
  \end{equation*}
\item and finally, compute the gradient
  \begin{equation*}
    \nabla_\theta E(\theta) = \nabla_\theta I - (\partial_\theta u_0)^\top \mu(0) - \int_0^T (\partial_\theta b)^\top \mu \,dt\,.
  \end{equation*}
\end{enumerate}

A few comments are of note:
\begin{itemize}
\item In contrast to the sections before, in the current setup we are
  not considering the case of small noise. Instead, the fluctuations
  are held at fixed amplitude, but we consider events in the limit of
  infinite threshold, $z\to\infty$. It is in this limit that the LDP
  in~(\ref{eq:random-initial-LDP}) is obtained, which might lead to a
  non-standard large deviation speed as a consequence. Therefore, the
  LDT computation discussed in this section can truly be considered
  for an \emph{extreme event instanton}.
\item Again, the formulation in the form of an adjoint
  equation~(\ref{eq:random-initial-adjoint}) is often beneficial from
  an implementation perspective as well: Many software packages for
  complex systems contain the computation of the adjoint
  field. Therefore, the computation of the gradient can possibly be
  achieved in a black-box manner.
\end{itemize}

\subsection{Example: Optimal excitation of the Fitzhugh-Nagumo model}
\label{sec:exampl-optim-excit-fitzhugh-nagumo}

\begin{figure}
  \begin{center}
    \includegraphics[width=248pt]{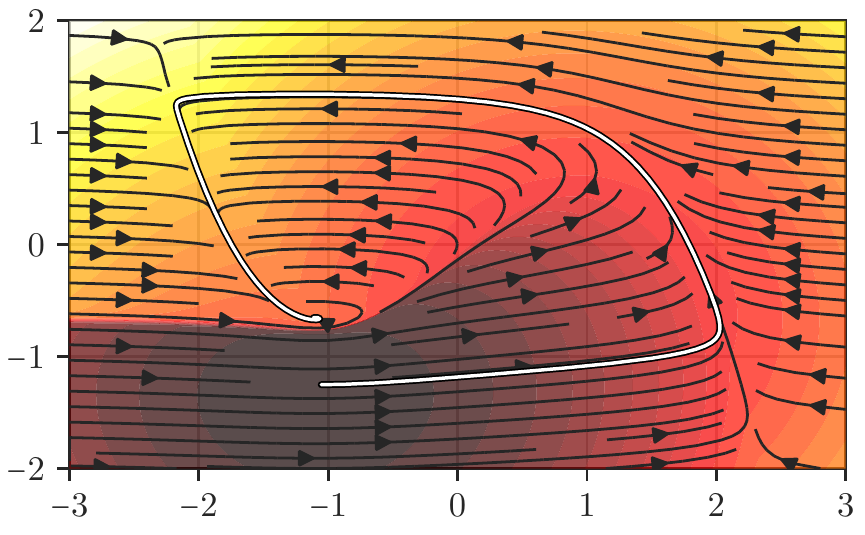}
  \end{center}
  \caption{Optimal perturbation of the initial condition to achieve an
    extreme excursion in the Fitzhugh-Nagumo
    model~(\ref{eq:fitzhugh-nagumo}). The flow field denotes the drift
    term, and the color denotes the value of the objective
    function. The trajectory realizing the maximal excursion is
    indicated as white line.\label{fig:fitzhugh-nagumo}}
\end{figure}

As an example, consider the following version of the deterministic
Fitzhugh-Nagumo model,
\begin{equation}
  \label{eq:fitzhugh-nagumo}
  \begin{cases}
    \dot x = \nu^{-1}(x-\tfrac13 x^3 - y)\\
    \dot y = x + a\,.
  \end{cases}
\end{equation}
If we consider the case $a>1$, this model is an excitable system, in
that there is a unique fixed point $(\bar x, \bar y) = (-a,\frac13
a^3-a)$, but small perturbations out of this fixed point potentially
lead to large excursions until the system returns to its steady
state. Here, we are interested in the optimal perturbation of the
initial condition away from the fixed point to achieve a large
excursion. For $\theta\in\Omega = \RR^2$, we define the distribution
of initial conditions as Gaussian centered around the fixed point,
\begin{equation*}
  (x_0(\theta),y_0(\theta)) = \theta \sim \exp\left(\tfrac1{2\Delta}\left((x_0-\bar x)^2 + (y_0-\bar y)^2\right)\right)\,,
\end{equation*}
and take as observable
\begin{equation}
  \label{eq:FM-observable}
  F(\theta) = \max_{t\in[0,T]} x(t,\theta)\,,
\end{equation}
i.e.~the maximal excursion in the $x$-component of the trajectory. We
want to know $P(z) = P(F(\theta)\ge z)$, and the corresponding most
likely initial conditions (and trajectory) that realize this extreme
event. Note that~(\ref{eq:FM-observable}) is an observable of the
third form of~(\ref{eq:random-initial-F}), and the algorithm lined out
above has to be modified slightly. In particular, for $u(t) =
(x(t),y(t))$ we obtain the gradient $\nabla_\theta E$ from forward and
backward equations, which are, respectively,
\begin{equation}
  \begin{cases}
    \partial_t u = b(u), &u(0) = (x_0, y_0)\\
    \partial_t \mu = -(\partial_u b)^\top\mu, &\mu(t^*) = \partial_u f(u(t^*)) = (1, 0)\,,
  \end{cases}
\end{equation}
where $t\in[0,t^*]$ and $t^*$ is the time at which the maximum is
reached. We are considering only the first local maximum in time, but
the dynamics of~(\ref{eq:fitzhugh-nagumo}) are such that the first
local maximum necessarily is the global maximum. In this situation,
there is no additional term coming from the dependence of $t^*$ on the
random parameters, since
\begin{equation*}
  \begin{aligned}
    \nabla_{\theta} f(u(t^*(\theta),\theta)) &= \partial_u f(u(t^*,\theta)) \partial_\theta u(t^*,\theta) \\&\ + \partial_u f(u(t^*,\theta)) \partial_t u(t^*) \partial_\theta t^*\,,
  \end{aligned}
\end{equation*}
and the second term disappears because at $t^*$ we have $\partial_t
u(t^*) = 0$. The gradient can then be computed as
\begin{equation*}
  \nabla_{\theta} E = \frac1\Delta (\theta - \bar u) - \lambda \mu(0)
\end{equation*}
for $u_0(\theta) = \theta \in \RR^2$ and $\bar u = (\bar x, \bar y)$.

Figure~\ref{fig:fitzhugh-nagumo} shows the result of the minimization
procedure: For $\Delta=2, \lambda=1, a=1.1, \nu=10^{-1}$, the shading
indicates the objective
function~(\ref{eq:objective-function-random-initial}) for every
initial condition. One can clearly make out the jump in the objective
function across the separatrix, where trajectories start exhibiting
large excursions. The trajectory starting at the minimum is the one
that maximizes the excursion in $x$-direction, before it decays to the
fixed point. The streamlines show the dynamics of the Fitzhugh-Nagumo
model~(\ref{eq:fitzhugh-nagumo}).

\section{Instantons as part of other rare event algorithms}
\label{sec:usage-inst-other}

While instantons as prototypical realizations of rare events can be
used for their own sake to estimate probabilities, relative stability,
and transition mechanisms, they can also be helpful as ingredient to
increase efficiency of other types of rare event algorithms. Most
notably, whenever rare events are sampled numerically by
\emph{tilting} a given stochastic process to facilitate a rare event
in an importance sampling setup, the instanton can be considered as
the \emph{optimal tilt}. In this section, we want to discuss how this
can be achieved in practice, and common pitfalls of this strategy:
First, in section~\ref{sec:inst-import-sampl}, we will show how to use
instantons to perform importance sampling for Monte-Carlo methods. In
section~\ref{sec:inst-clon-algor}, we use instantons to construct
weighting functions for genealogical particle algorithms. This will be
accompanied by an example computing the probability of infection rates
in a stochastic model for epidemiology.

\subsection{Instantons for importance sampling}
\label{sec:inst-import-sampl}

Consider an expectation of the form
\begin{equation}
  \label{eq:MC-expect}
  A^\epsilon= \EE \exp(\eps^{-1} F(X_T^\eps))
\end{equation}
for a random process $X^\eps_t\in\RR^d$, for example the one obeying
an SDE like~(\ref{eq:1}). We saw in
section~\ref{sec:rare-event-algor-1} how to compute the corresponding
instanton and get the dominating contribution in the limit
$\eps\to0$. In order to get hold of a proper quantitative estimate
of~(\ref{eq:MC-expect}), though, one would naively consider a Monte
Carlo method with estimator
\begin{equation}
  \label{eq:estimator-naive}
  \delta_\eps = \frac1M \sum_{i=1}^M \exp(\eps^{-1}F(X_T^{\eps,i}))\,,
\end{equation}
where $\{X^{\eps,i}_t\}_{i=1}^N$ are $N$ independent realizations of
the process.  This estimator is unbiased, meaning that
\begin{equation*}
  \EE \delta_\eps=A^\epsilon\,.
\end{equation*}
The relative error of this estimator,
\begin{equation}
  \label{eq:relative-error}
  e(\delta_\eps) =
  \frac{\text{std}(\delta_\eps)}{\text{mean}(\delta_\eps)}
  = \frac1{\sqrt{M}}\sqrt{\frac{\EE \exp(-2\eps^{-1}
      F(X_T^\eps))}{\big(\EE\exp(-\eps^{-1}  F(X_T^\eps))\big)^2}-1}
\end{equation}
describes the relative variance of the estimator. For example, for
$e(\delta_\eps)=1$, the typical fluctuations of the estimate are of
the size of the estimated value itself. The goal is to achieve a small
relative error. In practice, for rare events, one often struggles to
even keep $e(\delta_\eps)$ bounded for $\eps\to0$: Even though
increasing the number of samples improves the quality of the estimate,
$e(\delta_\eps)\to0$ for $M\to\infty$, the relative error increases
exponentially for fixed $M$ as $\eps\to0$. As a consequence,
estimating rare events with the naive
estimator~(\ref{eq:estimator-naive}) is impractical as the variance
blows up. The standard answer to this problem is to employ
\emph{importance sampling}, i.e.~introducing a new process $Y^\eps_t$
under which the rare event becomes typical, but accounting for this
change of probability measure by correcting with the proper
Girsanov-factor. Indeed, considering
\begin{equation}
  dY_t^\eps = (b(Y_t^\eps) + \sigma v(t,Y_t^\eps))\,dt + \sqrt{\eps}\sigma \,dW_t\,,
\end{equation}
for some function $v:[0,T]\times\RR^d\to\RR^d$, we can express the
expectation~\eqref{eq:MC-expect} as
\begin{equation}
  \label{eq:Aeps2}
    A^\eps= \EE^x\exp(-\eps^{-1}
    F(Y_T^\eps) )M^\eps_T\,,
  \end{equation}
  where
  \begin{equation}
    M^\eps_T\!=\!\exp\Big(- \frac1{\sqrt{\eps}} \int_0^T \!\!\langle
    v(s,Y_s^\eps),dW_s\rangle
    - \frac1{2\eps}\int_0^T \!\!|v(s,Y_s^\eps)|^2\,ds\Big).
\end{equation}
This identity can be used to construct an unbiased estimator of
$A^\epsilon$ by replacing the expectation in~\eqref{eq:Aeps2} by an
empirical expectation over $M$ independent copies of $Y^\eps_t$, similar
to what was done to obtain the vanilla
estimator~\eqref{eq:estimator-naive}.  The question is how to best
choose the importance sampling bias~$v(t,x)$ to lower the variance of
this new estimator.  An intuitive idea would be to use the instanton
to do so~\cite{bucklew:2004, pelissetto-ricci-tersenghi:2014}. For
example, it has been suggested to take
\begin{equation}
  \label{eq:Cramerbias}
  \begin{aligned}
    v(t,x) &= \sigma^{-1}(\dot \phi(t) - b(x))\qquad \text{or}\\
    v(t,x) &= \sigma^T \theta(t)\,,
  \end{aligned}
\end{equation}
where $(\phi(t),\theta(t))$ is the instanton position and momentum
corresponding to the expectation~(\ref{eq:MC-expect}), i.e.~taking
$Y_t^\eps$ to be, respectively, the stochastic process
\begin{equation}
  \label{eq:importance-sampling-biased-process}
  \begin{aligned}
    dY_t^\eps &= \dot\phi(t)\,dt + \sqrt{\eps}\sigma\,dW_t \qquad
    \text{or}\\
    dY_t^\eps &= b(Y_t^\eps)\,dt + a \theta(t)\, dt + \sqrt{\eps}\sigma\,dW_t\,.
  \end{aligned}
\end{equation}
The intuition is that using either one of the processes
in~\eqref{eq:importance-sampling-biased-process} biases the dynamics
towards the dominating path $\phi(t)$. Similar ideas, inspired from
lattice quantum chromodynamics, have entered through stochastic field
theory to bias Monte Carlo methods with the knowledge of the
instanton. For example
in~\cite{ebener-margazoglou-friedrich-etal:2018} the instanton for the
stochastic Burgers equation is used precisely in the way
of~(\ref{eq:importance-sampling-biased-process}) to sample a modified
process describing the fluctuations around it, getting improved
statistics in the rare event regime.

Although it has been pointed out that this strategy does not succeed
to decrease variance in general, or might even perform worse than the
naive one in the limit as $\epsilon \to0$~\cite{glasserman-wang:1997},
it can be modified~\cite{vanden-eijnden-weare:2012} to achieve optimal
variance decay by recomputing the instanton trajectory on-the-fly.

\subsection{Instantons for cloning algorithms}
\label{sec:inst-clon-algor}

There is another way to incorporate knowledge of the instanton within
importance sampling, namely through algorithms of genealogical
type~\cite{DelMoral:2004gw, cerou-guyader:2007,%
  giardina-kurchan-lecomte-etal:2011,wouters-bouchet:2016}. In these
methods, an ensemble of trajectories (aka particles, copies, or
clones) is integrated, and particles are removed or duplicated
according to some rating that selects behaviors favorable to the event
at hand. To explain how this can be done in the context of rare event
algorithms, let us focus on the second choice for $v(t,x)$ in
\eqref{eq:Cramerbias} since out of the two it is the one that requires
the least modification of the drift\footnote{In complex situations,
  such as applications in meteorology or climate where only a
  block-box type solver is available, explicit modification of the
  drift should be kept at a minimum.}. The second equation
in~\eqref{eq:importance-sampling-biased-process} reads
\begin{equation}
  \label{eq:importance-sampling-biased-processb}
  dY_t^\eps = b(Y_t^\eps)\,dt + a \theta(t)\, dt + \sqrt{\eps}\sigma\,dW_t
\end{equation}
along with the estimator for~\eqref{eq:MC-expect}
\begin{equation}
  \label{eq:Aeps3}
  \begin{aligned}
    A^\eps = \EE^x\exp\Big(&-\eps^{-1}
    F(Y_T^\eps) - \frac1{\sqrt{\eps}} \int_0^T \langle
    \theta(t),\sigma dW_t\rangle
    \\&- \frac1{2\eps}\int_0^T \langle \theta(t), a \theta(t) \rangle\,dt\Big).
  \end{aligned}
\end{equation}

We begin by rewriting this last formula in a form that is more
convenient for resampling. To this end, let us integrate the following
identity
\begin{equation}
  \label{eq:7}
  \begin{aligned}
    d \langle \theta(t), Y^\eps_t\!-\!\phi(t)\rangle &= \langle
    \dot \theta(t), Y^\eps_t\!-\!\phi(t)\rangle dt + \langle
    \theta(t), b(Y^\eps_t) \!-\! b(\phi(t))\rangle dt\\
    &\ \ + \sqrt{\eps} \langle \theta(t), \sigma dW_t\rangle
  \end{aligned}
\end{equation}
to get
\begin{equation}
  \label{eq:8}
  \begin{aligned}
    -\frac1{\sqrt{\eps}} \int_0^T \langle \sigma^T\theta(t),
    dW_t\rangle &= -\frac1\eps \langle \theta(T), Y^\eps_T
    -\phi(T)\rangle\\&\ \  + \frac1\eps \int_0^T \alpha(t,Y^\eps_t)\,dt\,,
  \end{aligned}
\end{equation}
where we defined
\begin{equation}
  \label{eq:9}
  \alpha(t,x) = \langle \dot \theta(t), x - \phi(t)\rangle +
  \langle\theta(t), b(x)- b(\phi(t))\rangle\,.
\end{equation}
These manipulations allow us to write the
expectation~\eqref{eq:Aeps3} as
\begin{equation}
  \label{eq:10}
  \begin{aligned}
    A^\eps = \EE^x \exp\Big(&-\eps^{-1}
    F(Y_T^\eps) -\eps^{-1} \langle \theta(T), Y^\eps_T
    -\phi(T) \rangle\\&-\tfrac12\eps^{-1} \int_0^T \langle \theta(t),
    a\theta(t)\rangle dt \Big) W^\eps_T\,,
  \end{aligned}
\end{equation}
where
\begin{equation}
  \label{eq:11}
  W^\eps_t = \exp\Big(\frac1\eps \int_0^t \alpha(s,Y^\eps_s) ds\Big).
\end{equation}

The expression~\eqref{eq:10} can be used to design a genealogical
algorithm in which $W^\eps_t$ is viewed as a weight that each of the particles
carries and according to which they are periodically resampled.  More
concretely, consider $M$ copies of the system, all evolving according
to the SDE~\eqref{eq:importance-sampling-biased-processb}. Denote by
$Y_t^{i}$ the position of the $i$-th copy at time $t$ and by $W_t^i$
its weight. The particle positions and  weights are evolved independently
on intervals $(t_{k-1},t_k)$, where $t_k$, $k\in \NN_0$ with $t_{k-1}
< t_{k}$ are \emph{selection steps} when the resampling occurs. It
proceeds as follows: Denoting by $\Delta W^i_k$ the weight accumulated
by particle $i$ on the interval $(t_{k-1},t_k)$, i.e.
\begin{equation}
  \label{eq:12}
  \Delta W^i_k = \exp\Big(\frac1\eps \int_{t_{k-1}}^{t_k} \alpha(s,Y^i_s) ds\Big)\,,
\end{equation}
we compute
\begin{equation}
  p^i_k= \frac{\Delta W^i_k}{\sum_{j=1}^N \Delta W^j_k}\,,\quad \Delta \bar W_k = \frac1M
  \sum_{i=1}^M \Delta W^i_k\,,
\end{equation}
and choose independently (with replacement) $M$ copies in the set
$\{Y^i_{t_k}\}_{i=1}^M$, using probability $p^i_k$ to pick copy
$Y^i_{t_k}$. We then use the resulting copies as new set
$\{Y^i_{t_k}\}_{i=1}^M$, assign to each the same weight
$W^i_{t_k} =\prod_{l=1}^k \Delta\bar W_l$, and repeat on the next
interval $(t_{k},t_{k+1})$.

As a result of this procedure, we have at any time a set of $M$ copies
with nearly uniform weights (since they only diverge from one another
during the intervals $(t_k,t_{k+1})$), that provides us with the
following expression for $A^\eps$ (compare with~\eqref{eq:10})
\begin{equation}
  \label{eq:cloning-tilted-estimator}
  \begin{aligned}
    A^\eps &= \exp\Big( -\tfrac12\eps^{-1} \int_0^T \langle \theta(t),
    a\theta(t)\rangle dt\Big) \EE^x \zeta_M \quad \text{with}\\
    \zeta_M &\!=\! \frac1M \sum_{i=1}^M \!\exp\!\Big(-\eps^{-1}
    F(Y_T^i) \!-\!\eps^{-1} \langle \theta(T), Y^i_T
    \!-\!\phi(T)\rangle \Big) W^i_T\,,
  \end{aligned}
\end{equation}
where $\EE^x$ denotes expectation over both the noise term
in~\eqref{eq:importance-sampling-biased-processb} and the resampling
steps.  If $M$ is large enough we can simply build an unbiased
estimator for $A^\eps$ by using $\zeta_M$ directly (i.e, removing the
expectation); or we can repeat the estimation $R$ times with $M$
copies and replace the expectation
in~\eqref{eq:cloning-tilted-estimator} with an empirical average over
the values of $\zeta_M$ calculated in these $R$ runs.

The variance of the estimator based
on~\eqref{eq:cloning-tilted-estimator} has been analyzed
e.g.~in~\cite{DelMoral:2004gw,cerou-guyader:2007}, where other
variants of the algorithm (e.g.~in terms of the resampling step)
are also discussed. Let us simply mention here that this estimator may
not be better behaved than the one directly based on~\eqref{eq:10}
(i.e.~the one using no resampling based on the values of the weights),
but it offers multiple possibilities of modifications that can
systematically improve its variance---we refer the reader
to~\cite{ferre-grafke-vandeneijnden:2019} for more details.

Similar approaches are possible with adaptive multilevel
splitting\cite{cerou-guyader:2007,
  brehier-gazeau-goudenege-etal:2016}, where again a rating function
(``reaction coordinate'') has to be found to evaluate the performance
of multiple copies, and for which the instanton dynamics can be taken
as input.

\subsection{Example: Epidemiology and vaccination at birth}

\begin{figure*}
  \begin{center}
    \includegraphics[width=468pt]{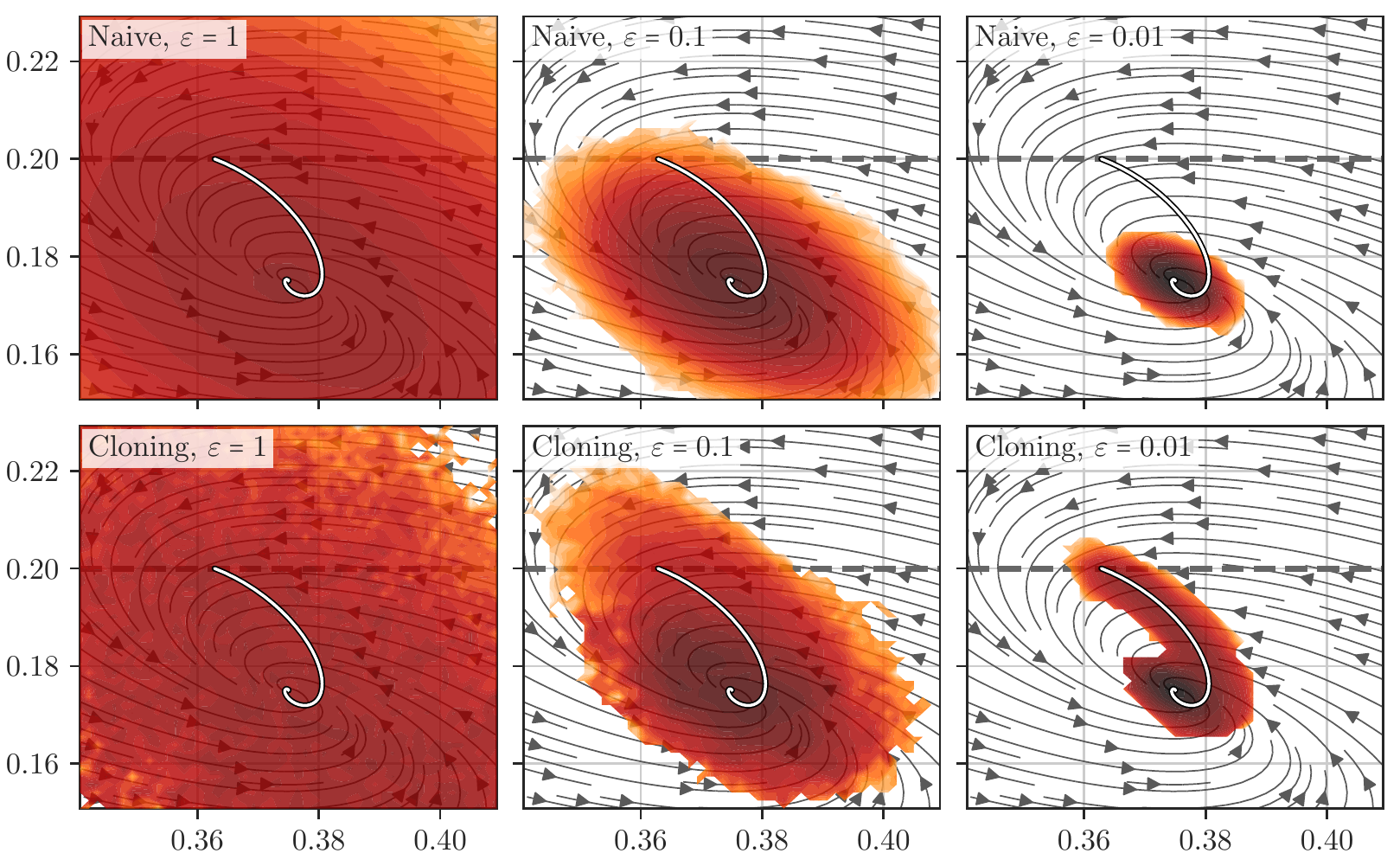}
  \end{center}
  \caption{Comparison of estimating the probability to reach an
    infection rate of 20\% with the naive and the cloning
    estimator. The heatmap depicts the logarithm of an $(S(t),I(t))$
    histogram, the streamlines represent the deterministic
    dynamics~(\ref{eq:vaccination}). The white line represents the
    instanton trajectory, and the dashed line the infection
    threshold. All measurements are obtained with 10000
    copies.\label{fig:cloning}}
\end{figure*}

To illustrate the scheme above, we consider the following
compartmental model inspired from epidemiology, where the spread of a
disease is modeled in the presence of vaccination. The total
population of individuals, denoted by $N$, is comprised of individuals
susceptible to the disease ($S$), individuals that are infected ($I$),
individuals that are recovered and thus immune ($R$) and individuals
that are vaccinated and thus immune ($V$). Individuals are born and
die with the same rate $\mu$, so that the total population remains
constant. In this simple rendition of disease spread with vaccination,
the vaccine is administered at birth, and with a vaccination rate of
$q$. In other words, a child is born vaccinated with probability $q$
or susceptible otherwise. The disease is transmitted by contact
between infected and susceptible individuals, with contact rate
$\beta$, while recovery is associated with the recovery rate
$\gamma$. In total, the model therefore reads
\begin{equation}
  \label{eq:vaccination}
  \begin{cases}
    \dot S = \mu N (1-q) - \mu S - \beta N^{-1} IS\\
    \dot I = \beta N^{-1} IS - (\mu + \gamma) I\\
    \dot V = \mu N q - \mu V\\
    \dot R = \gamma I - \mu R\,.
  \end{cases}
\end{equation}
Interestingly, depending on the vaccination rate, the
model~(\ref{eq:vaccination}) results in either total eradication of
the disease after transient dynamics (disease free equilibrium), or a
fixed point where the disease is still present (endemic
equilibrium). More precisely, the reproduction number
\begin{equation*}
  R_0 = \frac{\beta}{\mu+\gamma}\,,
\end{equation*}
describes the average number of contacts per infected individual
(i.e.~the ratio between contact frequency and the frequencies
associated with recovery or death). If the vaccination rate $q$
exceeds a threshold $q^*$,
\begin{equation*}
  q\ge q^*= 1 - \frac1{R_0}\,,
\end{equation*}
then the disease will be eradicated eventually. Note that, since the
dynamics of $S$ and $I$ are independent of $V$ and $R$, it is enough
to consider the first two equations of (\ref{eq:vaccination}) to
establish whether the disease is eradicated in the long-time
limit. Furthermore, we will normalize the quantities to ratios in
$[0,1]$.

\begin{table*}
  \begin{tabular}{crrrrrrrrrr}
    $\varepsilon$        & 1.0    & 0.5    & 0.2    & 0.1    & 0.09   & 0.08   & 0.07   & 0.06    &  0.05   & 0.01
    \\[1ex]
    $e_{\text{naive}}$   & 0.0859 & 0.1399 & 0.4222 & 2.4839 & 3.3331 & 6.0061 & 9.4868 & 18.2391 & 31.6228 & ---\\
    $e_{\text{cloning}}$ & 0.1487 & 0.1590 & 0.1685 & 0.1729 & 0.1849 & 0.1816 & 0.1836 & 0.1928  &  0.1969 & 0.2749\\
  \end{tabular}
  \caption{Relative error of the naive estimator and the cloning
    estimator for different values of $\eps$. Each value is generated
    from 1000 experiments with 1000 copies each. The relative variance
    of the naive estimator diverges for
    $\eps\to0$.\label{tbl:variance}}
\end{table*}

In order to produce estimates of probabilities in this system, we
furthermore need to make assumptions about stochasticity present in
the quantities $S$ and $I$. It is natural to interpret the rate
equations~(\ref{eq:vaccination}) as the law of mass action of a
reaction network, transforming the species into each other, which
would lead to Poisson noise terms as encountered in
section~\ref{sec:gener-non-gauss}. For large population sizes, one
could also consider a multiplicative Gaussian noise, consistent with
the central limit theorem, similar to the discussion of the stochastic
Lotka-Volterra model in section~\ref{sec:exampl-lotka-volterra}. Using
this approximation, the stochastic system reads
\begin{equation}
  \begin{cases}
    dS &= \big(\mu N (1-q) - \mu S - \beta N^{-1} IS\big)\,dt\\
    &\ \ \  + \sqrt{\mu N (1-q) + \mu S + \beta N^{-1} IS}\,dW_S\\
    dI &= \big(\beta N^{-1} IS - (\mu + \gamma) I\big)\,dt\\
    &\ \ \ + \sqrt{\beta N^{-1} IS + (\mu + \gamma) I}\,dW_I\,,
  \end{cases}
\end{equation}
where $W_S$ and $W_I$ are independent Wiener processes.

As observable, we want to estimate the probability that after time $T$
we have reached an unusually high ratio $z\in[0,1]$ of infected
individuals, $P(I(T)\ge z)$, which we can write as expectation via
\begin{equation*}
  P(I(T)\ge z) = \EE \Theta(I(T)-z)\,,
\end{equation*}
for the Heaviside step function $\Theta$. Using this observable, we
can compare the naive estimator with the cloning estimator.

We are choosing parameters $\mu=0.1$, $\beta=0.8$, $\gamma=0.2$ and
$N=1500$, which result in a critical vaccination rate of
$q^*=0.625$. Here, we set $q=0.1$ instead, resulting in an endemic
equilibrium $(\bar S, \bar I) = (0.375,0.175)$. The threshold is set
to $z=0.2$, i.e.~we want to estimate the probability to have a ration
of infected individuals above 20\% after $T=100$. The results for three
different values of $\eps$ are shown in figure~\ref{fig:cloning},
depicting the logarithm of a histogram of final trajectories
$(S(t),I(t))$. The streamlines describe the deterministic drifts given
in equation~(\ref{eq:vaccination}), while the white line is the
instanton to reach the threshold. Indeed, we observe that for
$\eps=1$, reaching the threshold infected rate is not a rare event,
the instanton has no predictive power and the cloning algorithm
performs equally or worse to naive sampling. For $\eps=10^{-1}$, the
event is rarely observed for naive sampling, resulting in larger
variance estimates of the probability. The configurations resulting
from the cloning algorithm instead show a higher prevalence of
increased infection rates. This effect is especially pronounced for
$\eps=10^{-2}$, where we do not observe any sample reaching the
threshold in the naive case, and where the cloning samples clearly
track the instanton trajectory towards the threshold. The relative
errors of the two estimators are summarized in
table~\ref{tbl:variance}. With decreasing $\eps$ (and therefore
exponentially decreasing probability of the rare event), the variance
of the naive estimator blows up, while the variance of the cloning
estimator remains largely unchanged.

\section{Conclusion}
\label{sec:conclusion}

Summarizing, in this review we presented a collection of algorithms to
estimate rare event probabilities and properties by computing the
large deviation minimizer (instanton). They are largely divided into
two categories:

In the first category, one minimizes the rate function globally, by
discretizing it and then employing numerical minimization
techniques. Traditional members of this category are the minimum
action method (MAM)~\cite{e-ren-vanden-eijnden:2004} and the geometric
MAM (gMAM)~\cite{heymann-vanden-eijnden:2008}. Here, we provide a
simplified and optimized version of the second, the simplified gMAM,
that allows for carrying out the optimization in the space of
arc-length parametrized curves with a minimal number of necessary
derivatives of the large deviation Hamiltonian. Effectively, this
translates into gains in either run-time or implementation complexity
over traditional variants. Methods in this category are particularly
suited for computing transition trajectories between two sets or
points.

In the second category, one instead solves the Hamilton's equations
(or instanton equations) associated with the large deviation
Hamiltonian. In this category are the Chernykh-Stepanov
algorithm~\cite{chernykh-stepanov:2001} and its geometric
variant~\cite{grafke-grauer-schaefer-etal:2014}. Here, we provide an
interpretation of these algorithms in form of the adjoint formulation
of the optimization problem. Methods in this category are effectively
employed when the intention is to compute expectations along sample
paths, or loosely speaking most likely realizations of extreme events.

Even though these formalism constitute dual approaches to the same
problem, we saw that they are drastically different in terms of
applicability: For example, degenerate forcing is easily incorporated
into Hamilton's equations, but constitutes a numerical difficulty in
the form of stiff constraints for MAM-type algorithms. Conversely,
traversing a saddle point or crossing a separatrix is readily achieved
in MAM-type schemes, but leads to loss of convergence in the equations
of motion formulation.

Nevertheless, both approaches can be generalized to treat SDEs driven
by multiplicative noise as well as stochastic processes driven by
non-Gaussian noise. They can also handle, at least formally, infinite
dimensional processes, like the solutions of SPDEs. These approaches
can also be extended in multiple ways. Here we discussed how related
considerations apply to the case of dynamical systems with generic
random parameters and we also showed how instantons can be used as
input in importance sampling algorithms.

\begin{acknowledgments}
  We thank Freddy Bouchet, Gr\'egoire Ferr\'e, Robert Jack, and
  Jonathan Weare for useful discussions about genealogical
  algorithms. EVE is supported in part by the Materials Research
  Science and Engineering Center (MRSEC) program of the National
  Science Foundation (NSF) under award number DMR-1420073 and by NSF
  under award number DMS-1522767.
\end{acknowledgments}

\end{document}